\documentclass[11pt,a4paper]{article}
\usepackage{tikz}
\usepackage[compat=1.0.0]{tikz-feynman}
\usepackage{authblk}
\usepackage{placeins}
\usepackage{epsfig}
\usepackage{subfigure}
\usepackage{cancel}
\usepackage{comment}
\usepackage{amsmath}
\usepackage{amssymb}
\usepackage{ulem}
\usepackage{mathtools}
\usepackage{graphicx}
\usepackage{MnSymbol}
\graphicspath{{plots/}}
\usepackage{hyperref}
\usepackage{csquotes}
\usepackage{multirow}
\usepackage[left=.45in,right=.45in,top=.6in,bottom=.6in,headheight=14.5pt]{geometry}
\usepackage{multirow}
\usepackage{colortbl}
\usepackage{xcolor}
\usepackage{pdflscape}
\usepackage{orcidlink}
\usepackage[title]{appendix}
\usepackage[left=.45in,right=.45in,top=.6in,bottom=.6in,headheight=14.5pt]{geometry}
\usepackage{fmtcount} 
\usepackage{array,multirow}
\newcolumntype{C}[1]{>{\centering\arraybackslash}m{#1}}
\usepackage{epsfig}
\usepackage{float}
\usepackage{geometry}
\usepackage{amsmath}
\usepackage{amssymb,tikz-feynman}
\usepackage{cite}
\usepackage{ctable}
\newgeometry{vmargin={10mm,20mm},hmargin={19mm,19mm}}
\title{\textbf{\textcolor{blue}{Dark Matter and Collider Phenomenology in Radiative Type-III Seesaw Model with Two Inert Doublets}}}

\author{Tapender\thanks{tapenderphy@gmail.com}}
\author{Labh Singh\thanks{sainilabh5@gmail.com}}
\author {Surender Verma\thanks{s\_7verma@hpcu.ac.in}}
\date{}
\affil{Department of Physics and Astronomical Science, Central University of Himachal Pradesh, Dharamshala-176215, INDIA}
\begin{document}	
\maketitle
\begin{abstract}
\noindent We investigate a minimal Type-III scotogenic model featuring two inert scalar doublets and a hyperchargeless triplet fermion. The scalar sector, in addition to the Standard Model Higgs, includes a rich spectrum of dark scalars comprising two CP-even, two CP-odd, and two charged states. This framework gives rise to two viable dark matter candidates: the lightest CP-even dark scalar and the neutral component of the triplet fermion. We perform a comprehensive analysis of both dark matter scenarios, carefully examining their viability under the umbrella of theoretical consistency conditions and experimental constraints. Beyond the conventional collider signatures anticipated in the Type-III scotogenic model with a single inert doublet, our extended framework predicts distinctive and novel signatures.  
\end{abstract}
	
\section{Introduction} 
\noindent Two of the most profound mysteries in the Standard Model (SM) of particle physics are the origin of neutrino masses and the nature of dark matter. Neutrino oscillation experiments have firmly established that neutrinos have small but nonzero masses \cite{Super-Kamiokande:1998kpq,SNO:2001kpb,SNO:2002tuh, SNO:2011hxd}, challenging the SM framework where neutrinos are massless. Similarly, the observed relic density of dark matter (DM), measured with remarkable precision by cosmic microwave background (CMB) experiments \cite{Planck:2018vyg}, indicates the existence of a stable, non-luminous component of the Universe that cannot be explained by SM particles alone. These compelling clues strongly suggest the need for new physics beyond the SM.

\noindent The Type-III scotogenic model, first proposed by Ma \cite{Ma:2008cu}, elegantly addresses the origin of neutrino masses and the nature of dark matter by introducing a discrete $Z_2$ symmetry and new particles: an inert scalar doublet ($\phi$) and three hyperchargeless ($Y=0$) triplet fermions ($\Sigma$) \cite{Chao:2012sz}. All these fields are odd under the $Z_2$ symmetry, which prevents neutrino masses at tree level and ensures the stability of the lightest $Z_2$-odd particle, making it a natural dark matter candidate. Neutrino masses arise from quantum loop corrections involving the inert scalar and triplet fermions, with their smallness explained by the heavy right-handed fermion scale and loop suppression, making the scotogenic model a compelling SM extension. Also, due to the $Z_2$ symmetry, the triplet fermion decouple from the neutrino sector and its neutral component could be a viable fermionic DM candidate which satisfies the relic density bound for $M_{\Sigma^{0}} \backsimeq 2.5$ TeV \cite{Chao:2012sz}. In this framework, achieving the correct relic density requires a relatively large Yukawa coupling in the term \( y \tilde{\phi}^\dagger \overline{\Sigma} L \), which can be realized by imposing a near-degeneracy between the CP-even and CP-odd components of the inert scalar, resulting in a highly suppressed quartic coupling \( \lambda_5 \sim 10^{-9} - 10^{-10} \) \cite{Ahriche:2017iar,Ahriche:2018ger}. Moreover, the scalar DM scenario in this model resembles the inert doublet model (IDM) \cite{Deshpande:1977rw,LopezHonorez:2006gr,Gustafsson:2010zz,Belyaev:2016lok,Ilnicka:2015jba,Jueid:2020rek,Basu:2020qoe,Kalinowski:2020rmb,Falaki:2023tyd,Ghosh:2024wxq,Ghosh:2021noq,Abouabid:2023cdz,Arhrib:2013ela,Eiteneuer:2017hoh,Kalinowski:2018ylg,Datta:2016nfz}, which is tightly constrained by DM direct detection experiments \cite{LZCollaboration:2024lux}, making the IDM highly fine-tuned. A possible resolution to this issue involves extending the scalar sector of the Type-III scotogenic model. In this context, several extensions of the minimal scotogenic model \cite{Ma:2006km,Toma:2013zsa,Borah:2020wut,Sarma:2020msa,Tapender:2024ktc} have been explored, where the scalar sector is augmented with additional singlet \cite{Beniwal:2020hjc,Farzan:2009ji,Ahriche:2016cio,Esch:2018ccs,Cohen:2011ec,Cheung:2013dua,DuttaBanik:2014iad}, inert doublet \cite{Ahriche:2022bpx,Hehn:2012kz,Fuentes-Martin:2019bue,Escribano:2020iqq,Garbrecht:2024bbo} and triplet scalar \cite{Hirsch:2013ola,Rocha-Moran:2016enp,Fiaschi:2018rky,Choubey:2017yyn,Batra:2022org,Singh:2023eye}. Also, due to its $SU(2)$ triplet nature, the triplet fermion in Type-III scotogenic model exhibits enhanced gauge interactions, leading to a higher production cross-section and distinct visible decay modes. This makes it a compelling candidate for experimental searches at colliders. Recent studies demonstrate that using fat-jet techniques, triplet fermions with high masses can be explored at LHC \cite{Ashanujjaman:2021zrh}.

\noindent We explore an extension of the Type-III scotogenic model by introducing an additional inert scalar doublet. Neutrino masses are generated at the one-loop level, which requires at least one triplet fermion, resulting in a framework with two inert doublets and a triplet fermion. In this setup, we examine both scalar and fermion dark matter scenarios. During the radiation-dominated era, a triplet fermion DM with a mass below 2.5 TeV tends to be under-abundant due to large annihilation and co-annihilation cross-sections into electroweak gauge bosons and fermions, resulting in a delayed freeze-out. However, the presence of a second inert doublet can significantly alter the dark matter phenomenology, potentially lowering the viable dark matter mass scale. Furthermore, we investigate the phenomenological implications of this model at both $e^{+}e^-$ and $pp$ colliders. The presence of an additional doublet and a triplet fermion could lead to distinctive signatures at high-energy experiments, including modified production cross-sections, decay channels, and final-state distributions. ATLAS \cite{ATLAS:2020wop} and CMS \cite{CMS:2019lwf} have established a lower bound on the mass of the triplet fermion at 790 GeV. At higher mass ranges, the production cross-section of the triplet fermion decreases significantly, making it challenging to probe effectively at the LHC. However, triplet fermions with masses as large as 1.5 TeV can still be explored, as shown in Ref. \cite{Ashanujjaman:2021zrh}. These collider signatures provide a promising avenue for testing the model and constraining its parameter space through future precision measurements.

The structure of this work is as follows: In Section 2, we present the extension of the Type-III scotogenic model. The theoretical and experimental constraints on the model are outlined in Section 3. Section 4 provides a detailed analysis of both scalar and fermion dark matter scenarios. The collider signatures associated with these dark matter candidates are examined in Section 5. Finally, concluding remarks are given in Section 6.

\section{Radiative Type-III Seesaw Model}{\label{s2}}
In our model, we want to study the dark matter (DM) and the origin of neutrino masses by considering a radiative Type-III seesaw model. In general, the minimum number of right-handed fermions required to get correct neutrino masses is two. However, we can make another setup in which we can just use only one right-handed fermion as shown in \cite{Hehn:2012kz,Fuentes-Martin:2019bue,Garbrecht:2024bbo}. For this, we need to add one more inert scalar to the particle content. This not only allows us to generate neutrino mass, but with the addition of another inert scalar, the DM analysis is also modified as new particles open up various new channels.
We have extended the Standard Model (SM) with two inert doublets $\phi_k$  $(k=1,2)$ and one fermionic triplet $\Sigma$. Abelian discrete $Z_2$ symmetry is used to separate the dark sector from the SM sector. Under this symmetry, all SM particles are even and all new particles are odd. The transformations under $Z_2$ symmetry can be written as:
	\[ \Sigma \xrightarrow{} -\Sigma,\,\,\, \phi_k\xrightarrow{}-\phi_k,\,\,\, \Phi\xrightarrow{}\Phi,\,\,\, \Psi_{\text{SM}}\xrightarrow{}\Psi_{\text{SM}},\]
	where $\Psi_{\text{SM}}$ denotes all the SM fermions and $\Phi$ denotes SM Higgs doublet.
	
	The relevant part of Yukawa Lagrangian can be written as
	\begin{equation}\label{lag1}
		-\mathcal{L}_Y= y_{ k ,\alpha} \tilde{\phi_k}^\dagger \overline{\Sigma} L_{\alpha}+\frac{M_{\Sigma}}{2}  \text{Tr}( \overline{\Sigma} \Sigma^c)+\text{H.c.}\,\,,
	\end{equation}
		where   $\alpha$= $e$, $\mu$, $\tau$ represent charged leptons, $k=1,2$ and  $M_{\Sigma}$ denote the mass of  fermion triplet. The symbol $L_{\alpha}$  denote left-handed lepton doublet and $y_{k,\alpha}$  denote Yukawa couplings of neutrinos and triplet fermion with inert scalar $\phi_k$. Also, $\tilde{\phi_k}=i \sigma_2 \phi_k^*$ where $\sigma_2$ is  Pauli spin matrix and $\Sigma^c=C\overline{\Sigma}^T$, $C$ is the charge conjugation matrix and fermion triplet, in $SU(2)$ representation, can be written as \cite{Chao:2012sz}
        \begin{equation}
            \Sigma= \begin{pmatrix}
                \frac{\Sigma^0}{\sqrt{2}} & \Sigma^+ \\
                \Sigma^- & -\frac{\Sigma^0}{\sqrt{2}}
            \end{pmatrix},
        \end{equation}
where $\Sigma^0$ and $\Sigma^\pm$ have  same mass $M_\Sigma$, at the tree level.\\
The scalar potential of our model under the exact $Z_2$ symmetry can be written  as  \cite{Ahriche:2022bpx}
	\begin{equation}\label{pot1}
		\begin{split}
			V(\Phi, \phi_k) = & \,\, -\mu^2_{\Phi}\Phi^{\dagger}\Phi+\frac{1}{6}\lambda_{\Phi}(\Phi^{\dagger}\Phi)^2+\mu^2_{k}\phi_k^{\dagger}\phi_k+\frac{1}{6}\lambda_k(\phi_k^{\dagger}\phi_k)^2  \\
			&+\kappa_k (\Phi^{\dagger} \Phi) (\phi_k^{\dagger} \phi_k )+\zeta_k (\phi_k^{\dagger}\Phi) (\Phi^{\dagger}\phi_k) +\omega_1 (\phi_1^{\dagger} \phi_1)(\phi_2^{\dagger} \phi_2)  + \omega_2 (\phi_2^{\dagger} \phi_1)(\phi_1^{\dagger} \phi_2) \\
			& +\big[\mu_3^2 \phi_1^{\dagger} \phi_2+\frac{1}{2}\eta_k (\Phi^{\dagger}\phi_k)^2+\eta_3 (\Phi^{\dagger}\phi_1)(\Phi^{\dagger}\phi_2)+\eta_4 (\phi_1^{\dagger}\Phi)(\Phi^{\dagger}\phi_2)+\text{H.c.}\big] \,,
		\end{split}
	\end{equation}
	where $k=1, 2$ and all scalar couplings  are  real parameters.\\
	The scalar doublets can be parameterized as,
	\begin{equation}
		\Phi =\begin{pmatrix}
			\xi^+\\
			\frac{1}{\sqrt{2}}(v+h+i \xi^0) \\
		\end{pmatrix},\;\;
		\phi_k =\begin{pmatrix}
			\chi_k^+\\
			\frac{1}{\sqrt{2}}(\chi_k^0+i S_k^0)  \\
		\end{pmatrix};\quad k=1,2
	\end{equation}
	where $v=246$ GeV is the vacuum expectation value ($vev$), $h$ is CP-even Higgs field, $\xi^+$ and $\xi^0$ are charged and CP-odd parts, respectively, of Higgs doublet. In inert sector we have $\chi^0_k$ and $S_k^0$ which are CP-even and CP-odd parts, respectively, and $\chi_k^+$ represent charged components of $\phi_k$.
	After electroweak symmetry breaking (EWSB), we have mass eigenstates which consists of  two CP-even states $H_{1,2}^0$, two CP-odd states $A_{1,2}^0$ and two pairs of charged scalars $H_{1,2}^\pm$. The mass eigenstates are related to gauge eigenstates as,
	\begin{equation}
		\begin{pmatrix}
			H_1^0\\
			H_2^0 \\
		\end{pmatrix} =\begin{pmatrix}
			c_H &s_H\\
			-s_H & c_H
		\end{pmatrix} \begin{pmatrix}
			\chi_1^0\\
			\chi_2^0 
		\end{pmatrix},\;\;
		\begin{pmatrix}
			A_1^0\\
			A_2^0 \\
		\end{pmatrix} =\begin{pmatrix}
			c_A &s_A\\
			-s_A & c_A
		\end{pmatrix} \begin{pmatrix}
			S_1^0\\
			S_2^0 
		\end{pmatrix},\;\;
		\begin{pmatrix}
			H_1^\pm\\
			H_2^\pm \\
		\end{pmatrix} =\begin{pmatrix}
			c_C &s_C\\
			-s_C & c_C
		\end{pmatrix} \begin{pmatrix}
			\chi_1^\pm\\
			\chi_2^\pm 
		\end{pmatrix},
	\end{equation}
	where $c_Y=\cos\theta_Y$, $s_Y=\sin\theta_Y$ with $\theta_H$, $\theta_A$ and $\theta_C$ are  mixing angles  for CP-even, CP-odd and charged scalars, respectively. In this work, we assume the hierarchy among the masses of inert scalars as $m_{H^0_1}<m_{A^0_1}<m_{H^\pm_1}<m_{H^0_2}<m_{A^0_2}<m_{H^\pm_2}$ to study dark matter phenomenology. \\
The scalar couplings can be written in terms of masses of inert scalars 
\begin{equation}
	\begin{aligned}
		\mu_1^2 &= m_{H_1^\pm}^2 c_C^2 + m_{H_2^\pm}^2 s_C^2 - \frac{1}{2} \kappa_1 v^2, \quad
		\mu_2^2 = m_{H_1^\pm}^2 s_C^2 + m_{H_2^\pm}^2 c_C^2  - \frac{1}{2} \kappa_2 v^2, \\
		\mu_3^2 &= c_C s_C \left( m_{H_2^\pm}^2 - m_{H_1^\pm}^2 \right), \\
		\zeta_1 &= \frac{1}{ v^2}\left(m_{H_1^0}^2 c_H^2 + m_{H_2^0}^2 s_H^2 + m_{A_1^0}^2 c_A^2 + m_{A_2^0}^2 s_A^2 - 2 \left( m_{H_1^\pm}^2 c_C^2 + m_{H_2^\pm}^2 s_C^2 \right)\right), \\
		\zeta_2  &=\frac{1}{ v^2}\left( m_{H_2^0}^2 c_H^2 + m_{H_1^0}^2 s_H^2 + m_{A_2^0}^2 c_A^2 + m_{A_1^0}^2 s_A^2 - 2 \left( m_{H_2^\pm}^2 c_C^2 + m_{H_1^\pm}^2 s_C^2 \right) \right), \\
		\eta_1  &= \frac{1}{ v^2}\left( \left( m_{H_1^0}^2 c_H^2 + m_{H_2^0}^2 s_H^2 \right) - \left( m_{A_1^0}^2 c_A^2 + m_{A_2^0}^2 s_A^2 \right) \right), \\
		\eta_2  &= \frac{1}{ v^2}\left( \left( m_{H_1^0}^2 s_H^2 + m_{H_2^0}^2 c_H^2 \right) - \left( m_{A_1^0}^2 s_A^2 + m_{A_2^0}^2 c_A^2 \right) \right), \\
		\eta_3  &= \frac{1}{ v^2}\left( s_H c_H \left( m_{H_2^0}^2 - m_{H_1^0}^2 \right) - s_A c_A \left( m_{A_2^0}^2 - m_{A_1^0}^2 \right) \right), \\
		\eta_4  &= \frac{1}{ v^2}\left( s_H c_H \left( m_{H_2^0}^2 - m_{H_1^0}^2 \right) + s_A c_A \left( m_{A_2^0}^2 - m_{A_1^0}^2 \right) - 2 c_C s_C \left( m_{H_2^\pm}^2 - m_{H_1^\pm}^2 \right) \right).
	\end{aligned}
\end{equation}
So, free parameters of our model are
$ \kappa_{1,2}, \; y_{k, \alpha}, \; m_{H_{1,2}^0}, \; m_{A_{1,2}^0},\; m_{H_{1,2}^\pm}, \; M_{\Sigma} \; \text{and} \; s_{H,A,C}$.

The Yukawa Lagrangian  part consisting of  inert scalar fields can be written, in terms of their mass eigenstates, as
\begin{equation}\label{lag_mass}
\begin{aligned}
	-\mathcal{L}_Y = & \sum_{\alpha, k} \big[ \frac{1}{2} \left( g_{k \alpha } H_k^0 + i f_{k \alpha } A_k^0 \right) \bar{\Sigma}^0 \nu_{\alpha L}  - h_{k \alpha} H_k^\pm  \bar{\Sigma}^+ \nu_{\alpha L}  \\ &+  \frac{1}{\sqrt{2}} \left( g_{k \alpha } H_k^0 + i f_{k \alpha } A_k^0 \right) \bar{\Sigma}^- l_{\alpha L} + \frac{1}{\sqrt{2}} h_{k \alpha} H_k^\pm  \bar{\Sigma}^0 l_{\alpha L}\big] + \text{H.c.},
 \end{aligned}
\end{equation}
$\text{with } \alpha = e, \mu, \tau \;\text{and}\; k = 1, 2 \;$
\begin{equation}
	\begin{aligned}
		 g_{1 \alpha} &= c_H y_{1, \alpha} + s_H y_{2,  \alpha}, \quad
		 g_{2 \alpha} = -s_H y_{1, \alpha} + c_H y_{2,  \alpha}, \\
		 f_{1 \alpha} &= c_A y_{1, \alpha} + s_A y_{2,  \alpha}, \quad
		 f_{2 \alpha} = -s_A y_{1, \alpha} + c_A y_{2,  \alpha}, \\
		 h_{1 \alpha} &= c_C y_{1, \alpha} + s_C y_{2,  \alpha}, \quad
		 h_{2 \alpha} = -s_C y_{1, \alpha} + c_C y_{2,  \alpha}.
	\end{aligned}
\end{equation}
The first two terms inside parenthesis in Eqn. (\ref{lag_mass}) induce neutrino mass at the one-loop level through the radiative seesaw mechanism. The third term, can contribute to the neutrino transition magnetic moment, while the last three terms play a role in lepton flavor violating decays.

After EWSB, the light neutrino mass generated at one-loop level can be expressed as 
\begin{equation}
	m_{\alpha \beta}^{(\nu)} =  \sum_{k=1}^2 M_{\Sigma} \left\{ g_{k \alpha } g_{k \beta } \mathcal{F} \left( \frac{m_{H_k^0}}{M_{\Sigma}} \right) - f_{k \alpha } f_{k \beta} \mathcal{F} \left( \frac{m_{A_k^0}}{M_{\Sigma}} \right) \right\},
\end{equation}


 with loop factor $\mathcal{F}(x) = \frac{x^2 \log(x)}{8\pi^2(x^2-1)}$.

 The neutrino mass matrix can be written as,
\begin{equation}
	(m^\nu)_{3 \times 3} = (y^T)_{3 \times 2} (\Lambda)_{2 \times 2} (y)_{2 \times 3},
\end{equation}
where $y$ is Yukawa coupling matrix and $\Lambda$ is a non-diagonal matrix which consists of  loop factors
\begin{equation}
	\Lambda = \begin{pmatrix}
		\Lambda_{11} & \Lambda_{12} \\
		\Lambda_{12}  &\Lambda_{22}
	\end{pmatrix},
\end{equation}

\begin{equation}
	\begin{aligned}
		\Lambda_{11} &=  M_{\Sigma} \left[ c_H^2 \mathcal{F} \left(\frac{m_{H^0_1}}{M_{\Sigma}}\right) +s_H^2 \mathcal{F} \left(\frac{m_{H^0_2}}{M_{\Sigma}} \right) - c_A^2 \mathcal{F} \left(\frac{m_{A^0_1}}{M_{\Sigma}} \right) - s_A^2 \mathcal{F} \left(  \frac{m_{A^0_2}}{M_{\Sigma}} \right) \right], \\
		\Lambda_{12} &=  M_{\Sigma} \left[ s_H^2 \mathcal{F} \left(\frac{m_{H^0_1}}{M_{\Sigma}}\right) +c_H^2 \mathcal{F} \left(\frac{m_{H^0_2}}{M_{\Sigma}} \right) - s_A^2 \mathcal{F} \left(\frac{m_{A^0_1}}{M_{\Sigma}} \right) - c_A^2 \mathcal{F} \left(  \frac{m_{A^0_2}}{M_{\Sigma}} \right) \right], \\
		\Lambda_{22} &=  M_{\Sigma} \left[ c_H s_H \left( \mathcal{F} \left(\frac{m_{H^0_1}}{M_{\Sigma}}\right) - \mathcal{F} \left(\frac{m_{H^0_2}}{M_{\Sigma}} \right)\right) - c_A s_A \left( \mathcal{F} \left(\frac{m_{A^0_1}}{M_{\Sigma}} \right) - \mathcal{F} \left(  \frac{m_{A^0_2}}{M_{\Sigma}} \right) \right) \right].
	\end{aligned}
\end{equation}
Using Casas-Ibarra parameterization \cite{Casas:2001sr}, we have Yukawa coupling matrix
\begin{equation}\label{eq_casas}
    y=U_{\Lambda}D_{\sqrt{\Lambda}}^{-1}RD_{\sqrt{m}}U^{\dagger},
\end{equation}
where $U_{\Lambda} \; \text{and}\; D_{\sqrt{\Lambda}}$ are diagonalizing and diagonal matrix for $\Lambda$, respectively. The  diagonal neutrino mass matrix  and Pontecorvo-Maki-Nakawaga-Sakata (PMNS) neutrino mixing matrix  are denoted by $D_{\sqrt{m}}\; \text{and}\; U$, respectively, and $R$ is orthogonal matrix with complex angle $\theta$. In this model, the lightest neutrino remains massless. We will consider normal hierarchy of neutrino masses, so $R$ can be written as,
 \begin{equation}
 	R = \begin{pmatrix}
 		0 & \cos\theta & \sin\theta \\
 		0& - \sin \theta   & \cos \theta
 	\end{pmatrix}.
 \end{equation}





\section{Constraints: Theoretical and Experimental}\label{sec3}
This model need to satisfy some theoretical and experimental constraints which we have discussed as follows:\\
\textbf{Theoretical constraints:} The couplings of the scalar potential need to satisfy the following theoretical constraints:
\begin{itemize}
	 \item \textbf{Vacuum Stability:} 
	 To ensure that in all directions of field space scalar potential is bounded from below we need scalar couplings  to satisfy some constraints \cite{Kannike:2012pe,Kannike:2016fmd,Kannike:2021fth,Ahriche:2015mea,Ahriche:2022bpx}:  $\lambda_\Phi>0$, $\lambda_1>0$, $\lambda_2>0$, $\lambda_{\Phi} \lambda_1 \lambda_2-\lambda_{\Phi} [\text{min}(0,\omega_1+\omega_2)]^2-\lambda_2 \text{min}(0,\kappa_1+\zeta_1+\eta_1)+2 \text{min}(0,\omega_1+\omega_2) \text{min}(0,\kappa_1+\zeta_1+\eta_1)\text{min}(0,\kappa_2+\zeta_2+\eta_3)-\lambda_{1}[\text{min}(0,\kappa_2+\zeta_2+\eta_2)]^2>0, \frac{8 \lambda_{\Phi} \lambda_{1} \lambda_2}{27}-\frac{2}{3}\lambda_2[\text{min}(0,\kappa_1+\zeta_1)]^2-\frac{2}{3} \lambda_{\Phi} [\text{min}(0,\omega_1+\omega_2)]^2+2\text{min}(0,\kappa_1+\zeta_1)\text{min}(0,\omega_1+\omega_2)\text{min}(0,\kappa_2+\zeta_2)-\frac{2}{3}\lambda_{1}[\text{min}(0,\kappa_2+\zeta_2)]^2>0$.
In addition to this, we also require,  $\mu_1^2>0,\; \mu_2^2>0, \mu_1^2+\mu_2^2-\sqrt{(\mu_2^2-\mu_1^2)^2+4(\mu_3^2)^2}>0$.  
	\item \textbf{Perturbativity:} For theory to remain in the perturbative regime we need scalar couplings to satisfy the following bounds:
	$\lambda_\Phi,\;\lambda_{1,2},\; |\kappa_{1,2}|,\; |\kappa_{1,2}+\zeta_{1,2}|,\;|\zeta_{1,2}+\kappa_{1,2}\pm \eta_{1,2}|,\;|\omega_1|,\;|\omega_1+\omega_2|,\;|\eta_{1,2}|,\;\frac{1}{2}|\zeta_{1,2}\pm\eta_{1,2}|,\;\frac{1}{2}|\omega_2|\leq4 \pi$. 
	 \item \textbf{Perturbative unitarity:}
	 All processes involving scalars or gauge bosons, perturbative unitarity need to be preserved. It is shown in Ref. \cite{Akeroyd:2000wc} that when eigenvalues ($\Lambda_i$) of scattering amplitude matrix  are smaller than $8\pi$ i.e., $|\Lambda_i|<8 \pi$  perturbative unitarity conditions are achieved.
	 In the models like we are investigating here, the full scattering amplitude matrix can be divided into six sub-matrices due to presence of some exact symmetries like CP, electric charge and global $Z_2$ symmetry. These sub-matrices and their basis are given in Ref. \cite{Ahriche:2022bpx}.	 
	 
\end{itemize}
\textbf{Experimental constraints:}
\begin{itemize}
	\item \textbf{Gauge bosons decay widths:} The decay widths of $W/Z$ bosons are measured with high precision at LEP and we do not want it to change due to decay of $W/Z$ bosons to $Z_2$ odd scalars. This can be achieved  if we assumes that $\text{min}(m_{H_1^\pm}+ m_{A_1^0},\; m_{H_1^\pm}+ m_{H_1^0})>m_W,\; \text{min}(m_{H_1^0}+ m_{A_1^0},\; 2 m_{H_1^\pm})>m_Z,\; M_{\Sigma^0}+M_{\Sigma^\pm}>m_W,\; 2 M_{\Sigma^\pm}>m_Z$. Here,  $m_W$ and $m_Z$ denote the masses of the $W$ and \( Z \) bosons, respectively, while $ M_{\Sigma^0}$ and  $M_{\Sigma^\pm}$ represent the masses of the neutral and charged components of the triplet fermion.

	\item \textbf{LEP direct searches of chargions and neutralinos:} The lower bounds on the masses of light neutral inert scalars ($H_1^0,\; A_1^0$) and charged inert scalar $H_1^\pm$ can be put by considering the null results of neutralinos and charginos searches at LEP \cite{DELPHI:2003uqw}. We have considered $\text{max}(m_{A_1^0},m_{H_1^0})>110$ GeV and $m_{H_1^\pm}>78$ GeV \cite{Lundstrom:2008ai,Pierce:2007ut}.
	\item \textbf{Lepton flavor violating (LFV) processes:}
   We have also considered the LFV bound for the  $\mu\rightarrow e \gamma$ process, which gives the most stringent upper bound for the branching ratio $Br(\mu \rightarrow e \gamma)<4.2\times10^{-13}$ \cite{MEG:2016leq}. The lepton flavor processes are induced at the one-loop level to which the triplet fermion along with the inert scalars gives  a contribution \cite{Chao:2012sz}.
	\item \textbf{Electroweak precision observables:}
   We require that the oblique parameters $S$, $T$ and $U$ remain within the following ranges: $S = -0.04 \pm 0.10$, $T = 0.01 \pm 0.12$ and $U = -0.01 \pm 0.09$, in order to maintain consistency with electroweak precision data \cite{ParticleDataGroup:2024cfk}.
 \item \textbf{Neutrino oscillation data:}
 Neutrino oscillation data within the 3$\sigma$ range \cite{Esteban:2024eli} is used for the calculation of Yukawa couplings using the Casas-Ibarra parameterization (see Eqn. \ref{eq_casas}). Thus, our model is consistent with neutrino oscillation data.
\end{itemize}

\section{Dark Matter Phenomenology}
 For the dark matter, we have two candidates in this model, \textit{i.e.}, either the neutral inert scalar $ H_1^0 $ or the neutral component of the triplet fermion $ \Sigma^0 $. Now, we will investigate the dark matter phenomenology of the model by considering both scenarios one by one. For this, we will make use of the SARAH toolchain. First, we have implemented the model in SARAH-4.15.2 \cite{Staub:2008uz,Staub:2013tta,Staub:2015kfa,Porod:2014xia} and generated the modules for SPheno-4.0.5 \cite{Porod:2003um,Porod:2011nf} and micrOMEGAs-5.3.41 \cite{Belanger:2001fz,Belanger:2020gnr,Belanger:2021smw,Alguero:2022inz,Belanger:2014vza}. SPheno-4.0.5 numerically calculates the mass spectrum, LFV observables, vertices, etc., while micrOMEGAs-5.3.41 computes the cold dark matter (CDM) relic density along with the spin-independent DM direct detection cross-section. In the following analysis, the free scalar couplings are kept within the range $0 \leq |\kappa_{1,2}| \leq 4\pi$, the mixing angles are varied such that $|s_{H,A,C}| \leq 1$ and Yukawa couplings $y_{i\alpha}$ with strength $\sqrt{4 \pi}> |y_{i\alpha}| \geq 10^{-4}$ are considered.

 \subsection{Scalar Dark Matter}
\begin{figure}[!htbp]
	\centering
	\begin{tabular}{cc} 
		\includegraphics[width=0.45\linewidth]{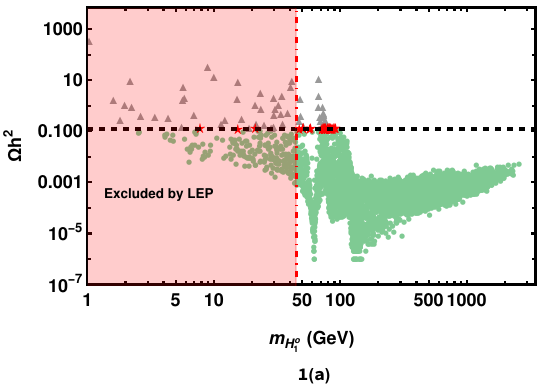}&\includegraphics[width=0.45\linewidth]{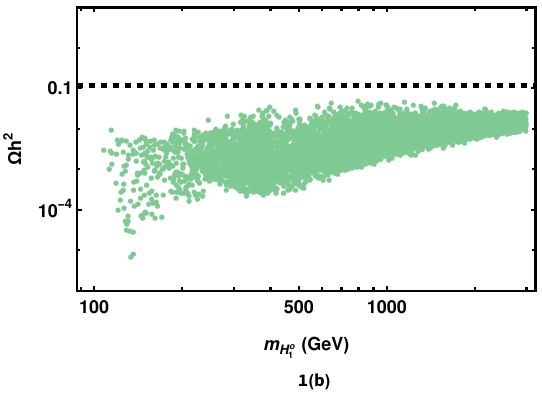}
	\end{tabular}
    \includegraphics[width=0.45\linewidth]{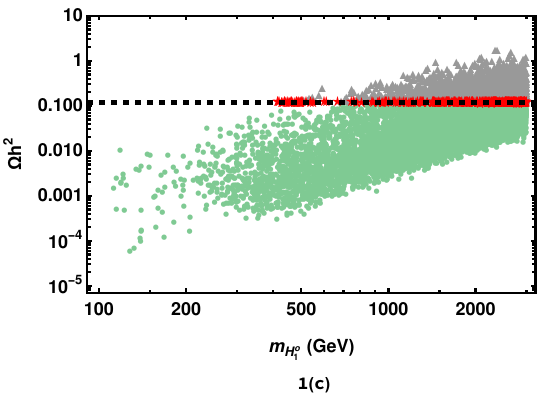}
	\caption{Figures \ref{fig1}(a) to \ref{fig1}(c) show the results obtained for CDM relic density for cases I to III, as listed in Table \ref{table2}, respectively. Light green points indicate an under abundant dark matter region, red stars correspond to values within the observed 3$\sigma$ range, and gray triangles represent an over abundant relic density. The black horizontal lines correspond to the experimental 3$\sigma$ range, while the red shaded region in Fig. \ref{fig1}(a) is excluded by LEP data.}
	\label{fig1}
\end{figure}

 We have studied dark matter phenomenology in the scalar sector by considering the neutral CP-even particle $H^0_1$ as dark matter (DM). As mentioned earlier, the hierarchy among the inert scalar particles is set to $m_{H^0_1}<m_{A^0_1}<m_{H^\pm_1}<m_{H^0_2}<m_{A^0_2}<m_{H^\pm_2}$. We have studied three cases  considering different  co-annihilation scenarios of DM  as shown in Table \ref{table2}\cite{Griest:1990kh}. In all cases, mass of triplet fermion is the largest. The mass of triplet fermion is varied considering lower bound of 790 GeV from ATLAS and CMS experiments. For higher mass region of DM due to perturbativity and unitarity conditions we need to have  small  mass splitting among inert scalar particles, so co-annihilation are unavoidable. However, to avoid co-annihilation in low mass region in different cases the mass range  for the particle next to  DM particle or its co-annihilating partner (CoP) is kept such that their mass splitting $(M_{Heavy}-M_{DM/CoP})/M_{DM/CoP} \geq 0.20$, hence there will be no unwanted co-annihilation upto 250 GeV mass of DM particle (see Table \ref{table2}). \\
 The relic density ($\Omega h^2$) of DM obtained for various cases considered in Table \ref{table2} is shown in Fig. \ref{fig1} as a function of DM mass. The light green points indicate under abundant DM relic density, red-star points lie within the observed 3$\sigma$ range \cite{Planck:2018vyg} and the gray-triangle points represent over abundant relic density. The black horizontal lines correspond to the experimental 3$\sigma$ range  of DM relic density \cite{Planck:2018vyg}.\\
 In case I only annihilation of $H^0_1$ is present upto 250 GeV as masses of other inert scalars and triplet fermion are higher (see Table \ref{table2}).  For this case DM mass versus the relic density ($\Omega h^2$) plot is shown in Fig. \ref{fig1}(a).

\begin{table}[!htbp]
    \centering
    \begin{tabular}{|c|c|c|c|}
        \hline
        Mass & Case I & Case II & Case III \\
        (GeV) &  &  &  \\
        \hline
        $m_{H^0_1}$  & $1 - 3000$  & $1 - 3000$  & $1 - 3000$  \\
        $m_{A_1^0}$  & $m_{H_1^0} + (50 - 150)$  & $m_{H_1^0} + (0.001 - 5)$  & $m_{H_1^0} + (0.001 - 5)$  \\
        $m_{H_1^\pm}$ & $m_{A_1^0} + (2 - 50)$  & $m_{A_1^0} + (0.001 - 5)$  & $m_{A_1^0} + (0.001 - 5)$  \\
        $m_{H_2^0}$  & $m_{H_1^\pm} + (2 - 50)$  & $m_{H_1^\pm} + (0.001 - 5)$  & $m_{H_1^\pm} + (0.001 - 5)$  \\
        $m_{A_2^0}$  & $m_{H_2^0} + (2 - 50)$  & $m_{H_2^0} + (0.001 - 5)$  & $m_{H_2^0} + (0.001 - 5)$  \\
        $m_{H_2^\pm}$ & $m_{A_2^0} + (2 - 50)$  & $m_{A_2^0} + (50 - 150)$  & $m_{A_2^0} + (0.001 - 5)$  \\
        $M_{\Sigma}$  & $m_{H_2^\pm} + (700 - 1500)$  & $m_{H_2^\pm} + (700 - 1500)$  & $m_{H_2^\pm} + (700 - 1500)$  \\
        \hline
    \end{tabular}
    \caption{Mass ranges for different dark sector particles in various cases for inert scalar dark matter.}
    \label{table2}
\end{table}

\begin{figure}[!htbp]
	\centering
	\includegraphics[width=0.45\linewidth]{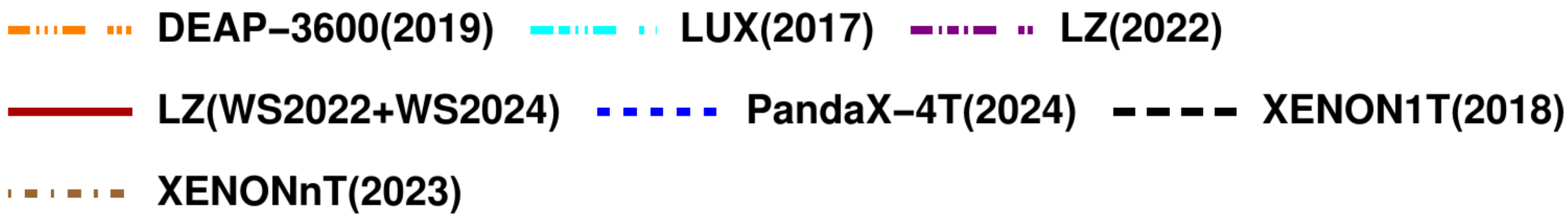}
	\begin{tabular}{cc} 
		\includegraphics[width=0.45\linewidth]{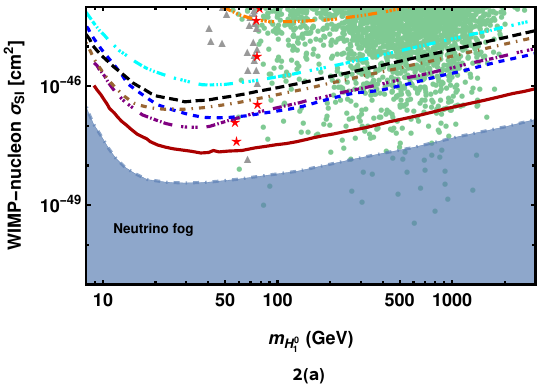}&\includegraphics[width=0.45\linewidth]{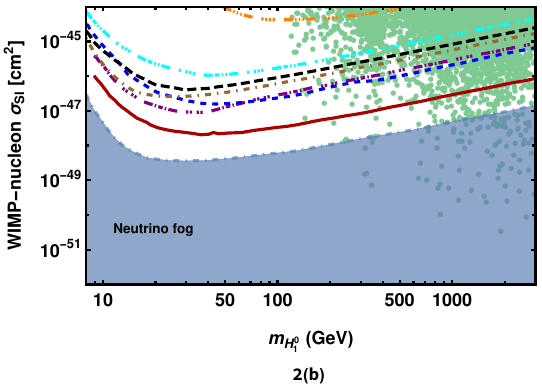}
	\end{tabular}
    \includegraphics[width=0.45\linewidth]{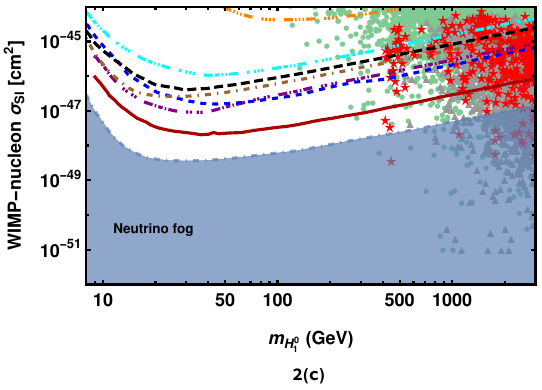} 
	\caption{The spin-independent nucleon–dark matter cross-section for cases I (2(a)), II (2(b)) and III (2(c)). The color coding is same as in Fig. \ref{fig1}.}
	\label{fig2}
\end{figure}\textbf{}
As evident from this figure,  in lower DM mass region upto the 100 GeV, the relic density can be under abundant, over abundant or lie within the experimental 3$\sigma$ region. However, region of DM mass upto 45 GeV is excluded by the LEP data, shown by the red shaded region, this sets a lower bound on the DM mass. We will only consider DM mass region which lies above it.  Around Higgs-resonance region, where DM mass is half of the mass of Higgs particle $m_h$ (\textit{i.e.}, $m_{DM}\approx$ $m_h/2\approx 62$ GeV), we have a sharp dip. This is due to the Higgs mediated $s-$ channel annihilation processes, in which DM+DM$\rightarrow h \rightarrow $SM+SM. The annihilation diagrams of type Figs. \ref{fig6}(a) and \ref{fig6}(b) dominantly contributes in this region (with $S$ being DM particle $H^0_1$). However, DM annihilation through  $t-$ channel processes mediated by charged components of triplet fermion $\Sigma^\pm$ to charged leptons  also gives contribution in this region shown in Fig. \ref{fig6}(c). In the region around the mass of W/Z boson ($m_{W^\pm}/m_{Z}$) i.e., $m_{H^0_1}=m_{W^\pm}=80$ GeV and $m_{H^0_1}=m_{Z}=92$ GeV, annihilation processes shown in Figs. \ref{fig6}(d) to \ref{fig6}(f) dominates with $W^\pm/Z$ in the final states. Another dip occurs as the mass of DM reaches around mass of Higgs approximately 125 GeV. In this region DM annihilation to Higgs particles dominates, the processes are shown in Figs. \ref{fig6}(g) and \ref{fig6}(h). 
 After this region processes where DM annihilates to Higgs particle $h$ and $W^\pm/Z$ bosons dominates, which continues in  higher mass region where co-annihilation processes such as shown in Fig. \ref{fig8}  can take place. So, in high mass region of DM all  processes shown from  Figs. \ref{fig6} to \ref{fig8}, giving Higgs particle $h$ and $W^\pm/Z$ bosons keep occurring which keeps the  relic density under abundant. These features are similar to the inert doublet model or Scotogenic model with one inert doublet. Here, relic density remains under abundant for higher DM mass.
The under abundant region obtained in this case is not disallowed but requires another partner to satisfy the observed relic density value.  Now to achieve the correct relic density at higher masses we will consider the co-annihilation cases as the presence of co-annihilation processes can increase or decrease the DM relic density by affecting the effective (co-)annihiation cross-section of DM.\\
So, in case II all inert scalars except for $H^\pm_2$  can co-annihilate starting from the lower DM mass. Now, because of experimental constraints on masses  as discussed in Section \ref{sec3}, lower mass of DM is not allowed, as evident from Fig. \ref{fig1}(b). From this figure we can see that, in all range of DM  mass relic density is under abundant. However, the relic density of DM is increased in the higher mass region in comparison to the case I. This implies that effective cross-section of DM (co-)annihilation is reduced. The dominating channels and processes remain the same as in case I with addition of co-annihilation processes. So, in higher mass region, main contribution comes from the DM to Higgs, $W^\pm$ and $Z$ (with their combinations like $hh$, $W^\pm W^\mp$, $ZZ$, $h W^+$ and $Z W^+$). The relevent annihilation and co-annihilation processes  are shown in Figs. \ref{fig6}-\ref{fig13}. \\
In case III,  all inert scalar particles are close in mass and thus can co-annihilate starting from lower DM mass region. Relic density and DM mass plot is shown in Fig. \ref{fig1}(c). The (co-)annihilation processes dominating  are similar to case II. 
As, evident from this figure that we have regions where relic density is under abundant, lying in $3\sigma$ range and over abundant. 
In this case,  we have region of DM mass starting around 414 GeV where DM relic density constraint can be satisfied (shown by  red-star points). This is one of the features of a model with two inert doublets \cite{Keus:2015xya}. In comparison to the inert doublet model or Scotogenic model with one inert doublet,  where DM relic density in the high mass region is satisfied above 500 GeV, here it can be as low as 414 GeV. This new region with low DM mass can have implications for searches of new particles in collider experiments. We will see this possibility later  in this paper.
  The impact of  decreasing the value of mass splitting among inert scalars can be seen in Figs. \ref{fig1}(b) and \ref{fig1}(c). More co-annihilating partners with appropriate value of mass splitting allow the relic density to reach near the experimentally observed region of CDM relic density. 
   All inert scalar particles can be almost degenerate with each other, this leads to the reduction of effective (co-)annihilation cross-section of DM. This reduction increases CDM relic abundance and we can even cross the experimentally observed region of CDM relic density.
    
The spin-independent (SI) WIMP-nucleon cross-section ($\sigma_{SI}$) as a function of the DM mass, $m_{H^0_1}$, for all cases listed in Table \ref{table2} are shown in Figs. \ref{fig2}(a) to \ref{fig2}(c), respectively. The upper limits from various dark matter direct detection experiments such as: combined  LUX-ZEPLIN (LZ) WIMP Search 2024 and 2022 (WS2024+WS2022) analysis \cite{LZCollaboration:2024lux}, LZ (2022) \cite{LZ:2022lsv}, LUX \cite{LUX:2016ggv}, PandaX-4T \cite{PandaX-4T:2021bab} (all power-constrained to $-1\sigma$), XENONnT \cite{XENON:2023cxc} (reinterpreted with a $-1\sigma$ power constraint), XENON1T \cite{XENON:2018voc} and DEAP-3600 \cite{DEAP:2019yzn} are, also, shown. The blue shaded area in the figure corresponds to ``neutrino fog" region \cite{OHare:2021utq}.\\
From these plots, it is evident that under the most stringent upper limit, which arises from the LZ WS2024+WS2022 analysis, the majority of parameter space is excluded. 
In case I for low DM mass region, as the Higgs dark matter coupling required in this region to satisfy the relic density is large which makes it difficult to reconcile it with SI WIMP-nucleon cross-section bound. However, for case III, in higher mass region a significant number of points which satisfies  experimental 3$\sigma$ bound on relic density of DM, remain within the allowed region.
 

\subsection{Fermionic Dark Matter}
We have studied the DM phenomenology of fermionic DM which in this model is the neutral component $\Sigma^0$ of fermion triplet $\Sigma$. At tree level all components of triplet fermion are mass degenerate. However, a small mass splitting arises between the neutral and charged components when  one loop correction to mass is considered \cite{Ma:2008cu}. This makes the $\Sigma^0$ slightly lighter among other components and thus makes it a DM candidate. We have studied two cases of DM  co-annhilation. For this study, we have followed the following mass hierarchy among the dark sector particles: $M_{\Sigma}<m_{H_1^0}<m_{A_1^0}<m_{H_1^\pm}<m_{H_2^0}<m_{A_2^0}<m_{H_2^\pm}$. Here, also, the mass of triplet fermion is varied keeping in mind the lower bound of 790 GeV from ATLAS and CMS experiments.

\begin{figure}[t]
	\centering
	\begin{tabular}{cc} 
		\includegraphics[width=0.45\linewidth]{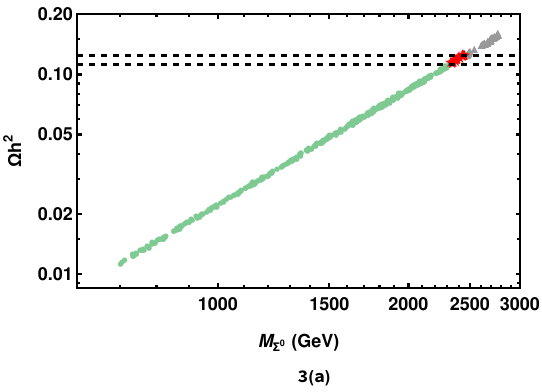} &  
		\includegraphics[width=0.45\linewidth]{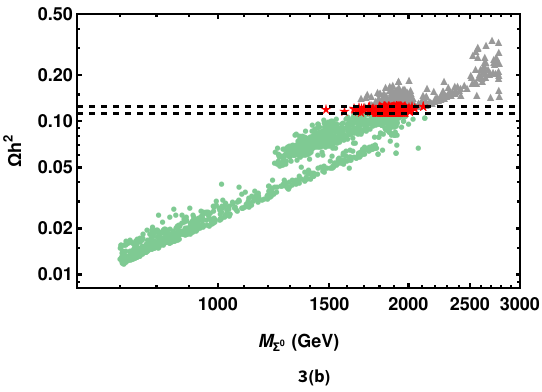}  
	\end{tabular}
	\caption{Figures \ref{fig4}(a) and \ref{fig4}(b) show the results obtained for CDM relic density for (co-)annihilation with fermion triplet components only and with all inert scalars, respectively. Light green points indicate an under abundant dark matter region, red stars correspond to values within the observed 3$\sigma$ range, and gray triangles represent an over abundant relic density. Black horizontal lines mark the experimental 3$\sigma$ range.}
	\label{fig4}
\end{figure}

In case I,  we have studied only (co-)annihilations among components of triplet fermion as mass of neutral component $\Sigma^0$ and charged components  $\Sigma^\pm$ are close to each other. However, by choosing the appropriate mass difference between lightest inert scalar  particle and fermion triplet we have avoided the co-annihilations of triplet fermions with inert scalars. For this purpose we have varied the mass of $H_1^0$ such that its mass splitting with triplet fermion is $\geq0.20$ i.e., $(m_{H_1^0}-M_\Sigma)/M_\Sigma \geq 0.20$ in whole range of DM mass. The other inert scalar particles  automatically satisfy this condition because of mass hierarchy we have chosen. The masses are varied as shown in second column of Table \ref{table3}. The results for DM relic density as a function of DM mass, after satisfying all the constraints mentioned in Section \ref{sec3}, are shown in Fig. \ref{fig4}(a). In  the plots shown in Fig. \ref{fig4}, the color and shape classification of the points has the same meaning as mentioned previously.
The various (co-)annihilation processes of DM  are shown in Fig. \ref{fig14}.


The DM relic density is  satisfied for region starting around 2332 GeV to 2452 GeV of DM mass. This obtained region for DM mass matches with the region mentioned in Ref. \cite{Ma:2008cu}, where SM is extended by a single triplet fermion without considering neutrino mass. 
Now we propose a scenario wherein  co-annihilation processes with inert scalars can lower the region of DM mass.

In case II, all inert scalar particles are closer in mass to the $M_{\Sigma}$ and can have co-annihilation with fermion triplet. The masses are varied as shown in third column of Table \ref{table3}. The result for this case is shown in Fig. \ref{fig4}(b). 
The various (co-)annihilation processes of DM relevant for this case are shown in Fig. \ref{fig14} and Fig. \ref{fig18}. Yukawa couplings will play an important role  as fermion triplet and inert scalars interaction depends on them. Large Yukawa couplings lead to the tension with the  experimental upper bound on BR$(\mu \rightarrow e \gamma)$.  The region where the relic density of DM lies within the experimental $3\sigma$ region now begins from a lower DM mass value, around 1483 GeV to 2112 GeV. This is possible due to additional co-annihilation processes of triplet fermion with inert scalar particles as shown in Fig. \ref{fig18}. This is remarkable, as in this case, the mass of the fermion triplet can be around 1483 GeV, which has significant implications for collider searches of the triplet fermion. This will be discussed in the next section. 


\begin{scriptsize}
\begin{table}[t]
    \centering
    \begin{tabular}{|c|c|c|}
        \hline
        Mass  & Case I & Case II\\
        (GeV) &  & \\

        \hline
        $M_{\Sigma}$  & $700 - 3000$& $700 - 3000$\\
        $m_{H_1^0}$  & $1.2 M_{\Sigma}$ to $(1.2 M_{\Sigma} + 100)$ & $M_{\Sigma} + (0.001 - 5)$\\
        $m_{A_1^0}$  & $m_{H_1^0} + (2 - 5)$& $m_{H_1^0} + (0.001 - 5)$\\
        $m_{H_1^\pm}$ & $m_{A_1^0} + (2 - 5)$& $m_{A_1^0} + (0.001 - 5)$\\
        $m_{H_2^0}$  & $m_{H_1^\pm} + (2 - 5)$& $m_{H_1^\pm} + (0.001 - 5)$\\
        $m_{A_2^0}$  & $m_{H_2^0} + (2 - 5)$& $m_{H_2^0} + (0.001 - 5)$\\
        $m_{H_2^\pm}$ & $m_{A_2^0} + (2 - 5)$& $m_{A_2^0} + (0.001 - 5)$\\
        \hline
    \end{tabular}
    \caption{Mass ranges for different dark sector particles in two cases for fermionic dark matter.}
    \label{table3}
\end{table}
\end{scriptsize}


\section{Collider Phenomenology}
 The model contains six inert scalars ($H_{1,2}^0,A_{1,2}^0,H_{1,2}^\pm$) along with  fermion triplet $\Sigma$ $(\Sigma^0,\Sigma^+,\Sigma^-)$. Their production at the collider gives us the opportunity to explore the possibility of detection in the collider experiments \cite{Ahriche:2017iar,Franceschini:2008pz,Belyaev:2016lok,Baumholzer:2019twf,Liu:2022byu,Ashanujjaman:2021zrh,Das:2020uer,Li:2022kkc,Singh:2025jtn,vonderPahlen:2016cbw,Datta:2016nfz,Kalinowski:2018kdn}.  
Here, we  focus on production of these particles in future linear $e^+e^-$ collider experiments like  CLIC (Compact Linear Collider) which will be designed for center-of-mass energies  380 GeV, 1.5 TeV and 3 TeV\cite{Aicheler:2018arh}, with integrated luminosities of 1 ab$^{-1}$, 2.5 ab$^{-1}$ and 5 ab$^{-1}$ respectively and ILC (International Linear Collider) designed for center-of-mass energy 500 GeV  with integrated luminosity of 4 ab$^{-1}$ \cite{Bambade:2019fyw}. We will also study the production of these particles at LHC (Large Hadron Collider) with center-of-mass energies 7 TeV and 14 TeV along with future hadron colliders:  HL-LHC (High Luminosity Large Hadron Collider) with center-of-mass energy 27 TeV\cite{Apollinari:2015bam} and FCC-hh (Future Circular Collider for hadron-hadron collision) with center-of-mass energy 100 TeV \cite{FCC:2018vvp}. To do the collider study we have used SARAH-4.15.2 \cite{Staub:2008uz,Staub:2013tta,Staub:2015kfa,Porod:2014xia} generated UFO model file \cite{Degrande:2011ua}, which then feed to the MadGraph5\_aMC@NLO version 3.5.4 \cite{Alwall:2014hca} to calculate the production cross-section of the various processes.\\ 
The production of particles at colliders depends on their mass. In this light, we explore the collider phenomenology of the newly obtained regions of DM mass in both inert scalar and fermion triplet DM scenarios. We have considered the two benchmark points BP1 and BP2 for $H_1^0$ as dark matter, as shown in second and third columns of Table \ref{tab5}, respectively. BP1 lies in the low DM mass region with DM particle mass $m_{H_1^0}\approx 60$ GeV, this point gives under abundant relic density. However, it contributes more than 50$\%$  to the total relic density. While BP2 lies in the newly obtained region of DM mass, with $m_{H_1^0} \approx 414$ GeV, which gives DM relic density within $3\sigma$ experimental range.  In the case where fermion triplet $\Sigma^0$ is taken as DM particle, we have considered one benchmark point BP3 which also lies in the newly obtained DM mass region, with triplet mass $M_{\Sigma^0}$ around $1483$ GeV, which satisfies relic density of DM at $3\sigma$ range, this point is shown in the fourth column of Table \ref{tab5}. All benchmark points (BP) satisfy the constraints considered in Section \ref{sec3}. In the following, we discuss in detail the collider phenomenology considering these benchmark points.

 \begin{table}[t]
	\centering
    \begin{scriptsize}
	\begin{tabular}{|c|c|c|c|} \hline 
		  Parameters&  BP1 &BP2 & BP3\\ \hline
		  $m_{H_{1,2}^0}$ (GeV)&  $(60.370,122.810)$&$(414.484, 415.982)$&$(1483.374, 1554.644)$\\ 
		  $m_{A_{1,2}^0}$ (GeV)&  $(111.724, 131.831)$&$(414.652, 416.354)$&$(1483.384,1554.645)$\\
		  $m_{H_{1,2}^\pm}$ (GeV)&  $(115.728, 135.695)$&$(415.248, 416.943)$&$(1483.536, 1554.981)$\\  
		  $M_{\Sigma^0}$ (GeV)&  $1205.384$&$1311.494$&$1482.741$\\
 $M_{\Sigma^\pm}$ (GeV)& $1205.384$& $1311.494$&$1482.789$\\
		  $s_{H,A,C}$&  $(-0.991, 0.943, -0.466)$&$(0.660, 0.572,-0.048)$&$(-0.379, -0.818,-0.187)$\\  
		  $\kappa_{1,2}$&  $(0.010, 0.438$)&$(-0.073, 0.163)$&$(0.712, -0.518)$\\
$y_{11}/ 10^{-3}$ & $1.762 + 0.576i$ & $0.666 - 2.875i$ & $8.479 - 7.572i$ \\
$y_{12}/ 10^{-3}$ & $-0.506 - 3.883i$ & $-4.213 + 1.710i$ & $-9.479 - 109.451i$ \\
$y_{13}/ 10^{-3}$ & $-3.247 - 3.217i$ & $-4.178 + 5.150i$ & $-65.597 - 72.573i$ \\
$y_{21}/ 10^{-3}$ & $0.711 - 0.346i$ & $-0.997 - 0.592i$ & $-55.304 + 7.470i$ \\
$y_{22}/ 10^{-3}$ & $-1.307 - 1.040i$ & $0.140 + 1.782i$ & $-262.601 + 471.215i$ \\
$y_{23}/ 10^{-3}$ & $-1.949 - 0.022i$ & $1.430 + 2.179i$ & $67.074 + 475.642i$ \\
		  $\Omega h^2$&  $0.090$&$0.121$&$0.119$\\
		  $Br(\mu \rightarrow e \gamma)$&  $3.4\times 10^{-21}$&$8.3\times10^{-20}$&$2.3\times10^{-13}$\\
		 $\sigma_{SI}$ (cm$^2$)& $8.18\times10^{-49}$&$5.19\times10^{-48}$&$-$\\ \hline
	\end{tabular}
	\caption{Benchmark points BP1 (second column) and BP2 (third column) correspond to  inert scalar ($m_{H_1^0}$) as dark matter  in the low  and  high  mass regions, respectively, while BP3 (fourth column) corresponds to fermion triplet ($\Sigma^0$) as dark matter. The elements of the Yukawa coupling matrix $y$ are denoted as $y_{ij}$, where $i = 1,2$ and $j = 1,2,3$.}
	\label{tab5}
  \end{scriptsize}
  \end{table}

 \subsection{Inert Scalars}
\begin{figure}[t]
	\centering
	\begin{tikzpicture}[scale=0.6]
		\begin{feynman}
			\vertex at (0,0) (i1);
			\vertex at (-1.5,0) (i2);
			\vertex at (1.5,1.5) (a);
			\vertex at (1.5,-1.5) (b);
			\vertex at (-2.5,1.5) (c);
			\vertex at (-2.5,-1.5) (d);
			
			\vertex at (-2.7,-1.7) () {\scriptsize\(e^-\)};
			\vertex at (-2.7,1.7) () {\scriptsize\(e^+\)};
			\vertex at (1.7,1.7) () {\scriptsize\(S^\prime\)};
			\vertex at (1.7,-1.7) () {\scriptsize\( S \)};
			\vertex at (-0.8,-0.2) () {\scriptsize\(h\)};
			\diagram*{
				(i2) -- [scalar] (i1), (i1) -- [scalar] (a), (b) -- [scalar] (i1), (i2) -- [] (c),(d) -- [] (i2)
			};
		\end{feynman}
		\node at (-0.6, -2.5) {(a)};
	\end{tikzpicture}
	\hspace{0.3cm}
	\begin{tikzpicture}[scale=0.6]
		\begin{feynman}
			\vertex at (0,0) (i1);
			\vertex at (-1.5,0) (i2);
			\vertex at (1.5,1.5) (a);
			\vertex at (1.5,-1.5) (b);
			\vertex at (-2.5,1.5) (c);
			\vertex at (-2.5,-1.5) (d);
			
			\vertex at (-2.7,-1.7) () {\scriptsize\(e^-\)};
			\vertex at (-2.7,1.7) () {\scriptsize\(e^+\)};
			\vertex at (1.7,1.7) () {\scriptsize\(S^\prime\)};
			\vertex at (1.7,-1.7) () {\scriptsize\( S \)};
			\vertex at (-0.8,-0.2) () {\scriptsize\(Z,\gamma\)};
			\diagram*{
				(i2) -- [boson] (i1), (i1) -- [scalar] (a), (b) -- [scalar] (i1), (i2) -- [] (c),(d) -- [] (i2)
			};
		\end{feynman}
		\node at (-0.6, -2.5) {(b)};
	\end{tikzpicture}
	\hspace{0.3cm}
	\begin{tikzpicture}[scale=0.6]
		\begin{feynman}
			\vertex at (0,0) (a);
			\vertex at (1.7,-1) (i1);
			\vertex at (-1.7,-1) (i2);
			\vertex at (0,1.5) (b);
			\vertex at (1.7,2.5) (c);
			\vertex at (-1.7,2.5) (d);
			\vertex at (1.8,2.7) () {\scriptsize\(S^\prime\)};
			\vertex at (-1.8,2.7) () {\scriptsize\(e^+\)};
			\vertex at (1.8,-1.3) () {\scriptsize\(S\)};
			\vertex at (-1.8,-1.3) () {\scriptsize\(e^-\)};
			\vertex at (1.0,1) () {\scriptsize\( \Sigma^0, \Sigma^- \)};
			
			\diagram*{
				(i1) -- [scalar] (a), (i2) -- [] (a),
				(b) -- [] (a), (b) -- [scalar] (c), (d) -- [] (b),
			};
		\end{feynman}
		\node at (0, -2.5) {(c)};
	\end{tikzpicture}
	\caption{The production modes of the inert scalars at the $e^+e^-$ collider where $S$ and $S^\prime$ belongs to inert scalars (for detail see Table \ref{tab6}).}
	\label{fig_inert_pro}
\end{figure}
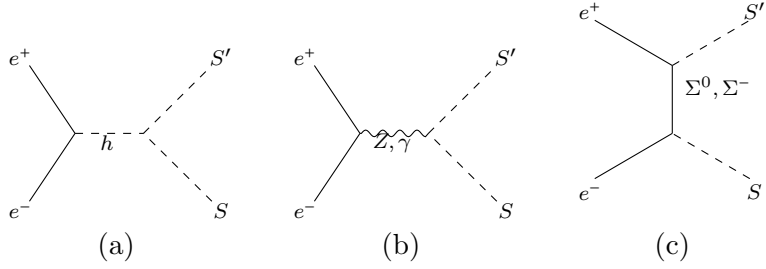

\begin{figure}[!htbp]
	\centering
	\begin{tabular}{cc} 
		\includegraphics[width=0.45\linewidth]{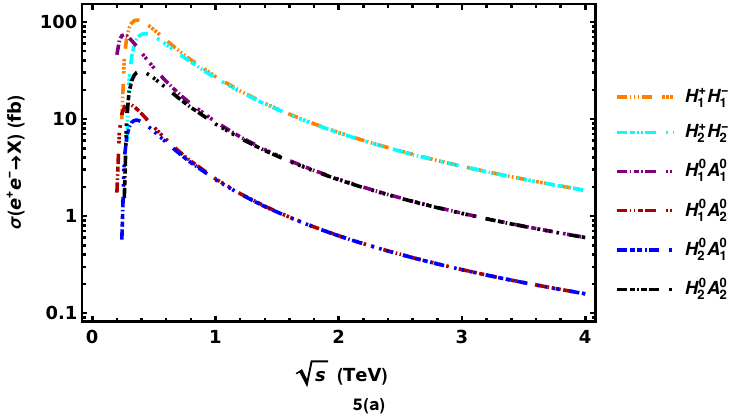}&\includegraphics[width=0.45\linewidth]{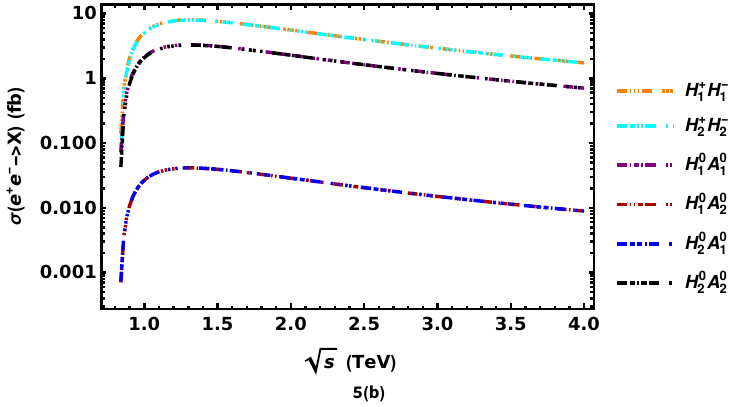}\\
	\end{tabular}
	\caption{Production cross-section for dominating processes with inert scalars in final state as a function of the center-of-mass energy $(\sqrt{s})$ in $e^+e^-$ collisions for BP1 (left) and BP2 (right) shown in Table \ref{tab5}.}
	\label{figee}
\end{figure}

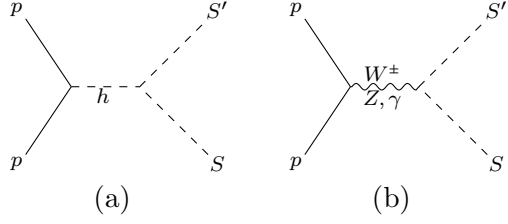
\begin{figure}[!htbp]
	\centering
	\begin{tikzpicture}[scale=0.6]
		\begin{feynman}
			\vertex at (0,0) (i1);
			\vertex at (-1.5,0) (i2);
			\vertex at (1.5,1.5) (a);
			\vertex at (1.5,-1.5) (b);
			\vertex at (-2.5,1.5) (c);
			\vertex at (-2.5,-1.5) (d);
			
			\vertex at (-2.7,-1.7) () {\scriptsize\(p\)};
			\vertex at (-2.7,1.7) () {\scriptsize\(p\)};
			\vertex at (1.7,1.7) () {\scriptsize\(S^\prime\)};
			\vertex at (1.7,-1.7) () {\scriptsize\( S \)};
			\vertex at (-0.8,-0.2) () {\scriptsize\(h\)};
			\diagram*{
				(i2) -- [scalar] (i1), (i1) -- [scalar] (a), (b) -- [scalar] (i1), (i2) -- [] (c),(d) -- [] (i2)
			};
		\end{feynman}
		\node at (-0.6, -2.5) {(a)};
	\end{tikzpicture}
	\hspace{0.3cm}
	\begin{tikzpicture}[scale=0.6]
		\begin{feynman}
			\vertex at (0,0) (i1);
			\vertex at (-1.5,0) (i2);
			\vertex at (1.5,1.5) (a);
			\vertex at (1.5,-1.5) (b);
			\vertex at (-2.5,1.5) (c);
			\vertex at (-2.5,-1.5) (d);
			
			\vertex at (-2.7,-1.7) () {\scriptsize\(p\)};
			\vertex at (-2.7,1.7) () {\scriptsize\(p\)};
			\vertex at (1.7,1.7) () {\scriptsize\(S^\prime\)};
			\vertex at (1.7,-1.7) () {\scriptsize\( S \)};
			\vertex at (-0.8,-0.3) () {\scriptsize\(Z,\gamma \)};
			\vertex at (-0.8,0.3) () {\scriptsize\( W^\pm \)};
			\diagram*{
				(i2) -- [boson] (i1), (i1) -- [scalar] (a), (b) -- [scalar] (i1), (i2) -- [] (c),(d) -- [] (i2)
			};
		\end{feynman}
		\node at (-0.6, -2.5) {(b)};
	\end{tikzpicture}
	\caption{The production modes of the inert scalars at the $pp$ collider where $S$ and $S^\prime$ belongs to inert scalars (for detail see Table \ref{tab7}).}
	\label{fig_pp_inert_pro}
\end{figure}

\begin{figure}[!hbtp]
	\centering
	\begin{tabular}{cc} 
		\includegraphics[width=0.45\linewidth]{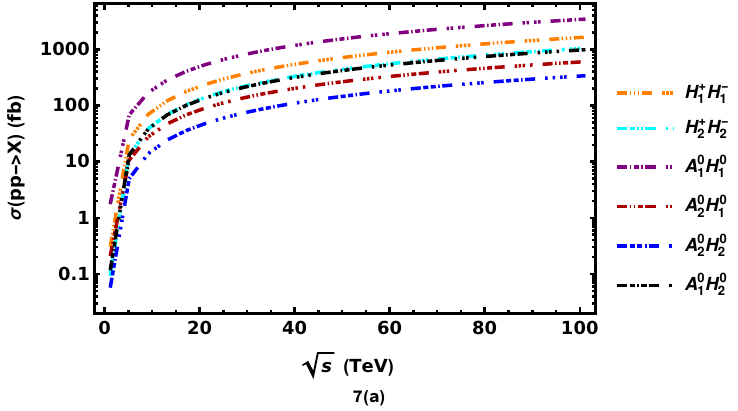}&\includegraphics[width=0.45\linewidth]{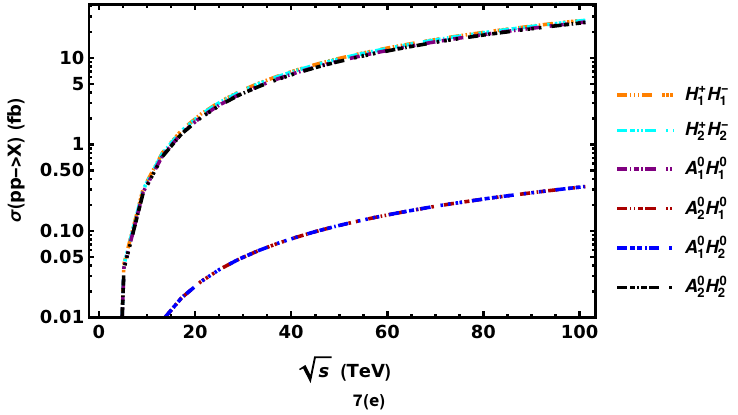}\\
		\includegraphics[width=0.45\linewidth]{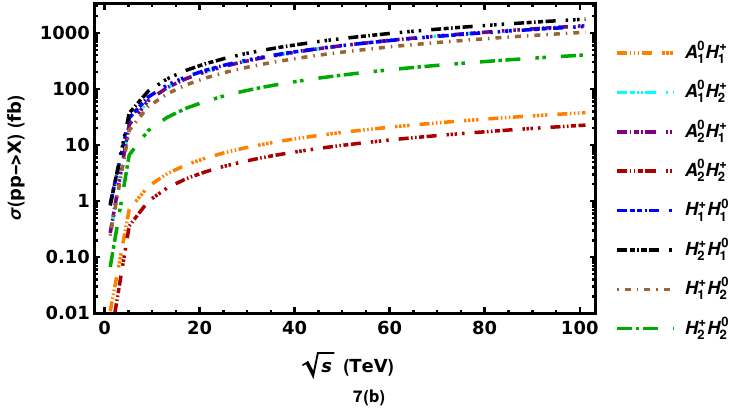}&\includegraphics[width=0.45\linewidth]{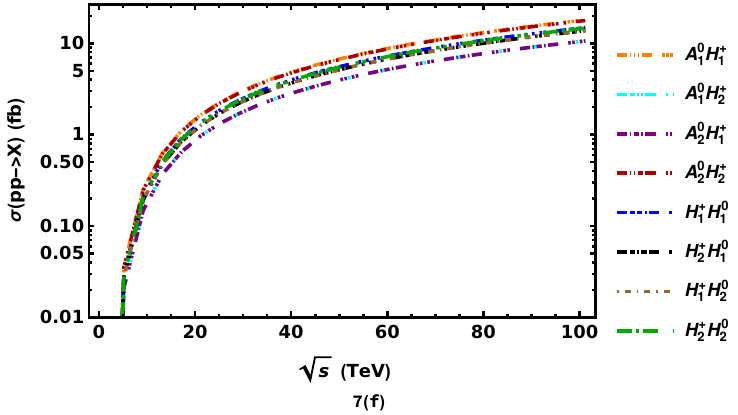}\\
		  \includegraphics[width=0.45\linewidth]{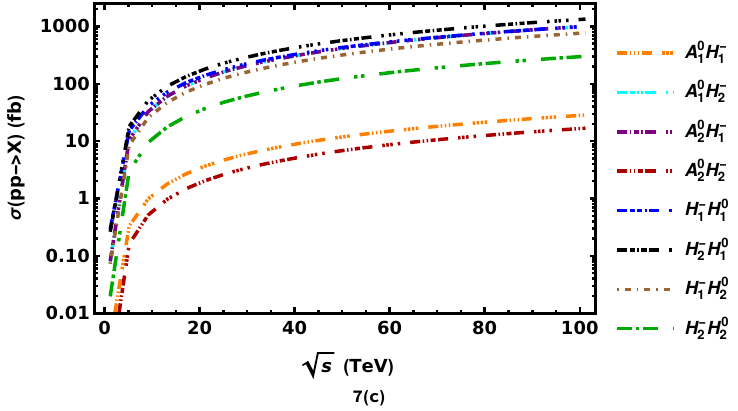}&\includegraphics[width=0.45\linewidth]{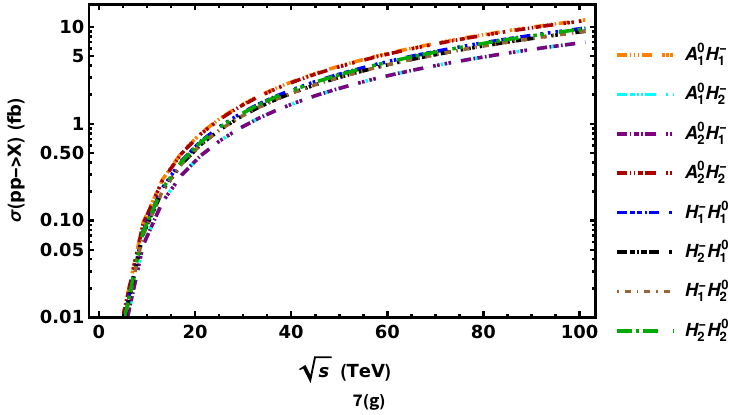}\\
		\includegraphics[width=0.45\linewidth]{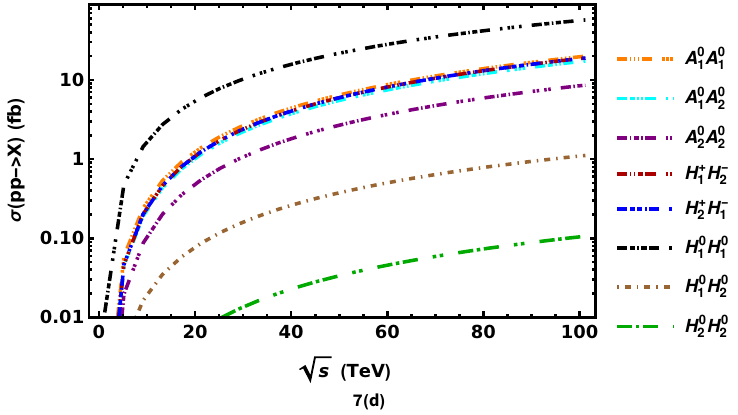}&\includegraphics[width=0.45\linewidth]{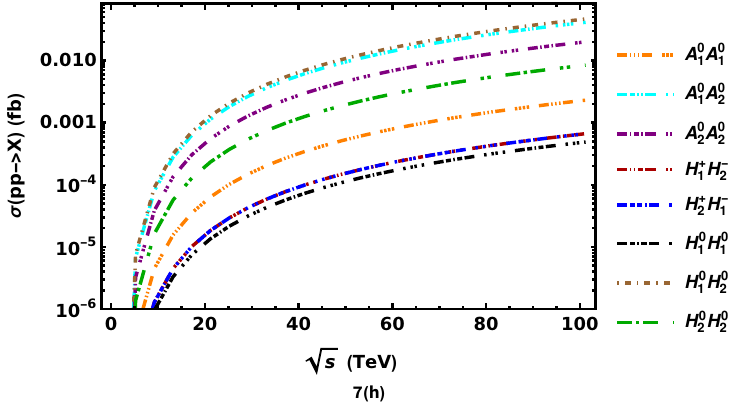}\\
	\end{tabular}
	\caption{Production cross-section for different inert scalars in final state as a function of the center-of-mass energy $(\sqrt{s})$ in $pp$ collisions for BP1 (left panel), BP2 (right panel) shown in Table \ref{tab5}.}
	\label{figpp}
\end{figure}

 As we mentioned earlier,  inert scalar sector is very rich in particle content. We have studied the prospects of production of these particles in future $e^+e^-$ collider experiments (CLIC and ILC) and at present and future hadron colliders (LHC, HL-LHC and FCC-hh). The inert scalars can be pair-produced in $e^+e^-$ collisions  which can consist of  charged inert scalars,  neutral inert scalars  or charged inert scalar in association with a neutral inert scalar. We have different processes for the production of inert scalars which consists of $s$-channel ($t$-channel) exchange of Higgs ($h$),  $Z$ bosons and photon ($\gamma$) ($\Sigma^0$ and $\Sigma^-$). The Feynman diagrams for different production modes of the inert scalars are shown in Fig. \ref{fig_inert_pro} and inert scalar pairs which can be produced in the final state are shown in Table \ref{tab6}.  By considering the BP1 and BP2 shown in Table \ref{tab5}, we obtained the production cross-section for the inert scalars as a function of center-of-mass energy ($\sqrt{s}$). The processes which does not involve $Z$  boson in  $s$-channel have very small production cross-section. So, we have investigated those $s$-channel  processes  which are mediated by  $Z$ boson, as shown in Table \ref{tab6}. The production cross-section for these particles in final states as a function of  $\sqrt{s}$ is shown in Fig. \ref{figee}(a) and \ref{figee}(b) for BP1 and BP2, respectively. The center-of-mass energy is varied  keeping in mind the physics potential of CLIC and ILC  experiments. The pair-production  cross-section  rises sharply after a certain threshold energy. This threshold energy corresponds to the sum of the mass of the produced particles, meaning production is not possible below this energy.
 As, $\sqrt{s}$ reaches threshold energy phase space for production opens up and cross-section increases sharply. After this peak, cross-section decreases because of the phase space suppression.  As evident from the Fig. \ref{figee}(a), for BP1 production cross-section for the  inert scalars pairs $H_1^+H_1^-$, $H_2^+H_2^-$ and $H_1^0 A_1^0$ dominates with  values around $105$ fb (at $\sqrt{s}=360$ GeV ), 76 fb (at $\sqrt{s}=440$ GeV ) and $74$ fb (at $\sqrt{s}=260$ GeV). Since, the masses for BP2 are high, production cross-section reduces and we get its highest value for $H_1^+ H_1^-$ and $H_2^+ H_2^-$  which is $8$ fb at  $\sqrt{s}=1.32$ TeV. \\
  We, also,  study the production of inert scalars at the LHC and future hadron collider energies HL-LHC and FCC-hh   for both benchmark points by varying  centre-of-mass energy upto $100$ TeV. The various production modes of inert scalar pairs in the final state at $pp$ collider is shown in Fig. \ref{fig_pp_inert_pro} and final states are listed in Table \ref{tab7}.  These production modes include $s$-channel processes mediated by Higgs ($h$), $W^\pm$ boson, $Z$ boson and photon ($\gamma$). The final states includes pairs of  neutral  or charged  inert scalars and charged inert scalar in association with neutral ones. We have shown the production cross-section for these inert scalar pairs as a function of $\sqrt{s}$ in left-panel (right-panel) of Fig. \ref{figpp} for BP1 (BP2). 
 The pair-production cross-section for inert scalars, mediated by $W^\pm$ and $Z$ bosons, is large in comparison to other processes. The  cross-sections for various the dominant pairs are shown in Table \ref{tab:dom_scalar} for BP1 and BP2 at various center-of-mass energies in current and future collider experiments.

\begin{table}[t]
	\centering
	\begin{tabular}{|c|c|} \hline 
		Mediator & Final state inert scalar pairs ($SS^\prime$)\\  \hline 
		Higgs ($h$) &  $H_1^+ H_1^-$, $H_2^+ H_2^-$, $H_1^+ H_2^-$, $H_2^+ H_1^-$, $A_1^0 H_1^0$, $A_2^0 H_1^0$,  $A_2^0 H_1^0$,  $A_2^0 H_2^0$,  $A_1^0 H_2^0$ \\
		&  $A_2^0 A_2^0$, $A_1^0 A_1^0$, $H_1^0 H_1^0$, $H_2^0 H_2^0$, $H_1^0 H_2^0$, $A_1^0 A_2^0$\\ \hline
		$Z$ boson & $H_1^+ H_1^-$, $H_2^+ H_2^-$, $A_1^0 H_1^0$, $A_2^0 H_1^0$, $A_2^0 H_2^0$, $A_1^0 H_2^0$\\ \hline
		Photon ($\gamma$)  &  $H_1^+ H_1^-$, $H_2^+ H_2^-$ \\ \hline 
		$\Sigma^0$  &   $H_1^- H_1^+$, $H_2^- H_2^+$, $H_2^- H_1^+$, $H_1^- H_2^+$  \\ \hline 
		$\Sigma^-$  & $H_1^0 A_1^0$, $H_1^0 A_2^0$, $A_2^0 H_2^0$, $A_1^0 H_2^0$, $A_1^0 H_1^0$, $A_2^0 H_1^0$, $A_2^0 H_2^0$, $A_1^0 H_2^0$  \\ 
		& $A_2^0 A_2^0$, $A_1^0 A_1^0$, $H_1^0 H_1^0$, $H_1^0 H_2^0$, $A_1^0 A_2^0$, $H_2^0 H_2^0$ \\ \hline
	\end{tabular}
	\caption{The inert scalar pairs in final state along with respective mediators in the e$^+$e$^-$ collision. The Feynman diagrams are shown in Fig. \ref{fig_inert_pro}.}
	\label{tab6}
\end{table}

\begin{table}[!htbp]
	\centering
	\begin{tabular}{|c|c|} \hline 
		Mediator & Final state inert scalar pairs ($SS^\prime$)\\  \hline 
		Higgs ($h$) &  $A_1^0 A_1^0$, $H_1^+ H_1^-$, $A_1^0 A_2^0$, $A_2^0 A_2^0$, $H_1^+ H_2^-$, $H_2^+ H_1^-$, $H_2^+ H_2^-$, $A_1^0 H_1^0$, $A_2^0 H_1^0$, $H_1^0 H_1^0$\\
		&  $H_1^0 H_2^0$, $A_1^0 H_2^0$, $A_2^0 H_2^0$, $H_2^0 H_2^0$ \\ \hline
		$W^\pm$ boson & $A_1^0 H_1^\pm$, $A_1^0 H_2^\pm$, $A_2^0 H_1^\pm$, $A_2^0 H_2^\pm$, $H_1^\pm H_1^0$, $H_2^\pm H_1^0$, $H_1^\pm H_2^0$, $H_2^\pm H_2^0$ \\ \hline
		$Z$ boson & $H_1^+ H_1^-$, $H_2^+ H_2^-$, $A_1^0 H_1^0$, $A_2^0 H_1^0$, $A_2^0 H_2^0$, $A_1^0 H_2^0$\\ \hline
		Photon ($\gamma$)  &  $H_1^+ H_1^-$, $H_2^+ H_2^-$ \\ \hline
	\end{tabular}
	\caption{The inert scalar pairs in final state along with their respective mediators in the $pp$ collision. The Feynman diagrams are shown in Fig. \ref{fig_pp_inert_pro}.}
	\label{tab7}
\end{table}

\begin{table}[h]
    \centering
    \begin{tabular}{lcccccccc}
        \toprule
       \textbf{Particle} & \multicolumn{2}{c}{\textbf{LHC 7 TeV}} & \multicolumn{2}{c}{\textbf{LHC 14 TeV}} & \multicolumn{2}{c}{\textbf{HL-LHC 27 TeV}} & \multicolumn{2}{c}{\textbf{FCC-hh 100 TeV}} \\
        \cmidrule(lr){2-3} \cmidrule(lr){4-5} \cmidrule(lr){6-7} \cmidrule(lr){8-9}
        \textbf{Pairs} & BP1 & BP2 & BP1 & BP2 & BP1 & BP2 & BP1 & BP2 \\
        \midrule
        $A_1^0H_1^0$  & 112  & 0.12  & 309  & 0.85  & 726  & 3    & 3404  & 25   \\
        $H_1^+H_1^-$  & 44   & 0.13  & 133  & 0.92  & 328  & 4    & 1625  & 27   \\
        $A_1^0H_2^0$  & 9    & -& 27   & -& 66   & -& 334   & -\\
        $H_2^+H_2^-$& 24   & 0.13      & 76   & 0.90     & 197  & 3    & 1034  & 27   \\
        $A_2^0H_1^0$  & 18   & 0.12  & 52   & 0.83  & 123  & 3    & 595   & 25\\
        $A_2^0H_2^0$& 24   & 0.12      & 75   & 0.83     & 189  & 3    & 972   & 25   \\
        $H_2^+H_1^0$  & 61   & -      & 164  & -     & 380  & -    & 1736  & -    \\
        $H_2^-H_1^0$& 32   & -      & 99   & -     & 253  & -    & 1321  & -    \\
        $H_1^+H_1^0$& 48   & 0.09      & 127  & 0.60     & 288  & 2    & 1287  & 15    \\
        $H_1^-H_1^0$& 26   & -      & 78   & -     & 195  & -    & 988   & -    \\
        $A_1^0H_2^+$& 40   & -      & 115   & -     & 276  & -    & 1325  & -    \\
        $A_1^0H_2^-$& 20   & -      & 67   & -     & 179  & -    & 991  & -    \\
        $A_2^0H_1^+$& 40   & -      & 115  & -     & 277  & -    & 1327  & -    \\
        $A_2^0 H_1^-$& 20   & -      & 67   & -     & 179  & -    & 992   & -    \\
        $H_1^+H_2^0$& 31   & -      & 89   & -     & 213  & -    & 1013  & -    \\
        $H_1^-H_2^0$& 16   & -      & 52   & -     & 138  & -    & 760   & -    \\
        $H_2^+H_2^0$& 12   & -      & 34   & -     & 82   & -    & 398   & -    \\
        $H_2^-H_2^0$& 6    & -      & 20   & -     & 53   & -    & 297   & -    \\
        $A_1^0H_1^+$& -    & 0.11      & -   & 0.72     & -   & 3    & -   & 18    \\
        $A_2^0H_2^+$& -    & 0.11      & -   & 0.70     & -   & 3    & -   & 17    \\
        \bottomrule
    \end{tabular}
    \caption{Dominating pair-production cross-section (fb) for inert scalars in $pp$ collision at different collider energies for BP1 and BP2. The symbol (-) signifies that the cross-section for the respective pair is small.}
    \label{tab:dom_scalar}
\end{table}

\subsection{Fermion Triplet}
\begin{figure}[t]
	\centering
	\begin{tikzpicture}[scale=0.6]
		\begin{feynman}
			\vertex at (0,0) (i1);
			\vertex at (-1.5,0) (i2);
			\vertex at (1.5,1.5) (a);
			\vertex at (1.5,-1.5) (b);
			\vertex at (-2.5,1.5) (c);
			\vertex at (-2.5,-1.5) (d);
			
			\vertex at (-2.7,-1.7) () {\scriptsize\(e^-\)};
			\vertex at (-2.7,1.7) () {\scriptsize\(e^+\)};
			\vertex at (1.7,1.7) () {\scriptsize\(\Sigma^+\)};
			\vertex at (1.7,-1.7) () {\scriptsize\( \Sigma^- \)};
			\vertex at (-0.8,-0.2) () {\scriptsize\(Z,\gamma\)};
			\diagram*{
				(i2) -- [boson] (i1), (i1) -- [] (a), (b) -- [] (i1), (i2) -- [] (c),(d) -- [] (i2)
			};
		\end{feynman}
		\node at (-0.6, -2.5) {(a)};
	\end{tikzpicture}
	\hspace{0.3cm}
	\begin{tikzpicture}[scale=0.6]
		\begin{feynman}
			\vertex at (0,0) (a);
			\vertex at (1.7,-1) (i1);
			\vertex at (-1.7,-1) (i2);
			\vertex at (0,1.5) (b);
			\vertex at (1.7,2.5) (c);
			\vertex at (-1.7,2.5) (d);
			\vertex at (1.8,2.7) () {\scriptsize\(\Sigma^+(\Sigma^0)\)};
			\vertex at (-1.8,2.7) () {\scriptsize\(e^+\)};
			\vertex at (1.8,-1.3) () {\scriptsize\(\Sigma^-(\Sigma^0)\)};
			\vertex at (-1.8,-1.3) () {\scriptsize\(e^-\)};
			\vertex at (-1.3,1) () {\scriptsize\( H_{1,2}^0, A_{1,2}^0 \)};
			\vertex at (1.3,1) () {\scriptsize\( (H_{1,2}^-) \)};
			
			\diagram*{
				(i1) -- [] (a), (i2) -- [] (a),
				(b) -- [scalar] (a), (b) -- [] (c), (d) -- [] (b),
			};
		\end{feynman}
		\node at (0, -2.5) {(b)};
	\end{tikzpicture}
		%
	\caption{The production modes of the $\Sigma^\pm \Sigma^\mp$ and $\Sigma^0 \Sigma^0$ at the $e^+e^-$ collider.}
	\label{fig_tri_pro}
\end{figure}
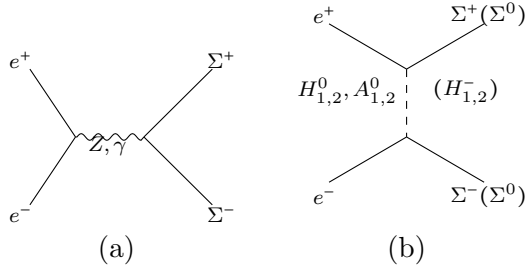

\begin{figure}[t]
	\centering
	\begin{tabular}{cc} 
	\includegraphics[width=0.45\linewidth]{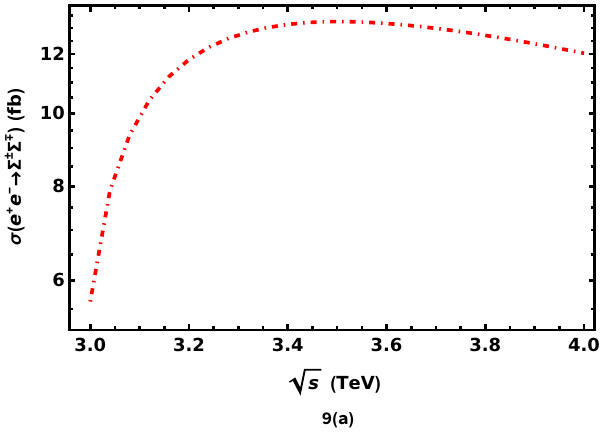}&\includegraphics[width=0.45\linewidth]{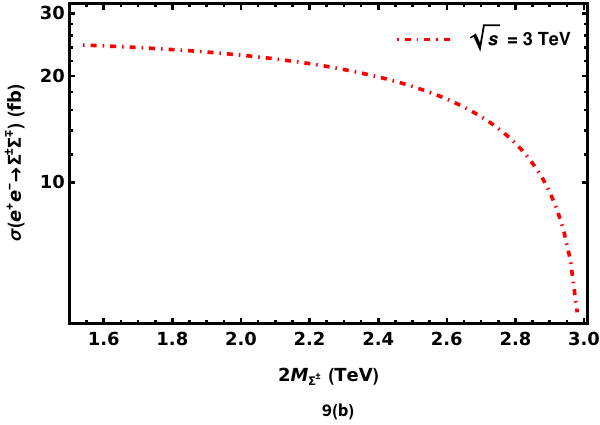}
	\end{tabular}
	\caption{Left panel: pair-production cross-section for $\Sigma^{\pm} \Sigma^{\mp}$ as a function of the center-of-mass energy $(\sqrt{s})$ in $e^+e^-$ collisions for benchmark point BP3, shown in Table \ref{tab5}.  Right panel: pair-production cross-section for $\Sigma^{\pm} \Sigma^{\mp}$  as a function of the triplet fermion pair mass   for CLIC center-of-mass energy ($\sqrt{s}$), 3 TeV.}
	\label{figtri}
\end{figure}

First, we  investigate the production of  triplet fermion  in  $e^+e^-$ collider experiments, such as CLIC and ILC. These particles can be pair-produced in   $e^+e^-$ colliders.  The production modes for $\Sigma^0$ pair include only  $t$-channel with charged inert scalars as mediators and $\Sigma^\pm$ are pair-produced via $s$-channel ($t$-channel) $Z$ boson and photon ($\gamma$)  (neutral inert scalars) as mediators as shown in Fig. \ref{fig_tri_pro}. The production cross-section for $\Sigma^\pm \Sigma^\mp$   pair  is shown in Fig. \ref{figtri}(a).
It is evident that  $\Sigma^\pm$, cross-section is maximum around $\sqrt{s}=3.52$ TeV with a value  around 13 fb. This lies outside the proposed center-of-mass energy for CLIC and ILC. However, at $\sqrt{s}=3$ TeV (CLIC center-of-mass energy) we have cross-section around $5$ fb. For $\Sigma^0\Sigma^0$ pair the production cross-section is very small hence, not shown here. This is because of the absence of  $s$-channel production mode. In Fig. \ref{figtri}(b), by considering the lower bound of $790$ GeV on triplet fermion mass given by ATLAS and CMS experiments, we have shown the production cross-section for $\Sigma^\pm\Sigma^\mp$ as a function of the triplet fermion pair mass, at CLIC center-of-mass energy,  $\sqrt{s}=3$ TeV. \\

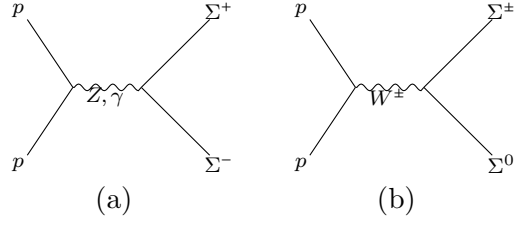
\begin{figure}[t]
	\centering
	\begin{tikzpicture}[scale=0.6]
		\begin{feynman}
			\vertex at (0,0) (i1);
			\vertex at (-1.5,0) (i2);
			\vertex at (1.5,1.5) (a);
			\vertex at (1.5,-1.5) (b);
			\vertex at (-2.5,1.5) (c);
			\vertex at (-2.5,-1.5) (d);
			
			\vertex at (-2.7,-1.7) () {\scriptsize\(p\)};
			\vertex at (-2.7,1.7) () {\scriptsize\(p\)};
			\vertex at (1.7,1.7) () {\scriptsize\(\Sigma^+\)};
			\vertex at (1.7,-1.7) () {\scriptsize\( \Sigma^- \)};
			\vertex at (-0.8,-0.2) () {\scriptsize\(Z,\gamma\)};
			\diagram*{
				(i2) -- [boson] (i1), (i1) -- [] (a), (b) -- [] (i1), (i2) -- [] (c),(d) -- [] (i2)
			};
		\end{feynman}
		\node at (-0.6, -2.5) {(a)};
	\end{tikzpicture}
	\hspace{0.3cm}
	\begin{tikzpicture}[scale=0.6]
		\begin{feynman}
			\vertex at (0,0) (i1);
			\vertex at (-1.5,0) (i2);
			\vertex at (1.5,1.5) (a);
			\vertex at (1.5,-1.5) (b);
			\vertex at (-2.5,1.5) (c);
			\vertex at (-2.5,-1.5) (d);
			
			\vertex at (-2.7,-1.7) () {\scriptsize\(p\)};
			\vertex at (-2.7,1.7) () {\scriptsize\(p\)};
			\vertex at (1.7,1.7) () {\scriptsize\(\Sigma^\pm\)};
			\vertex at (1.7,-1.7) () {\scriptsize\( \Sigma^0 \)};
			\vertex at (-0.8,-0.2) () {\scriptsize\(W^\pm\)};
			\diagram*{
				(i2) -- [boson] (i1), (i1) -- [] (a), (b) -- [] (i1), (i2) -- [] (c),(d) -- [] (i2)
			};
		\end{feynman}
		\node at (-0.6, -2.5) {(b)};
	\end{tikzpicture}
		%
	\caption{The production modes of the $\Sigma^\pm \Sigma^\mp$ and $\Sigma^0 \Sigma^\pm$ at the $pp$ collider.}
	\label{fig_pp_tri_pro}
\end{figure}

\begin{figure}[!htbp]
	\centering
    \includegraphics[width=0.45\linewidth]{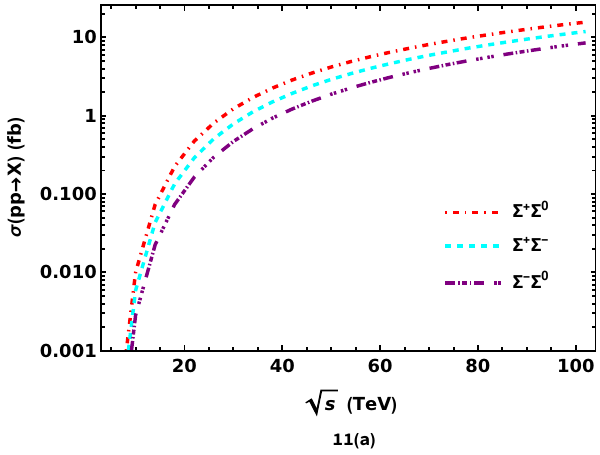}
	\begin{tabular}{cc} 
	\includegraphics[width=0.45\linewidth]{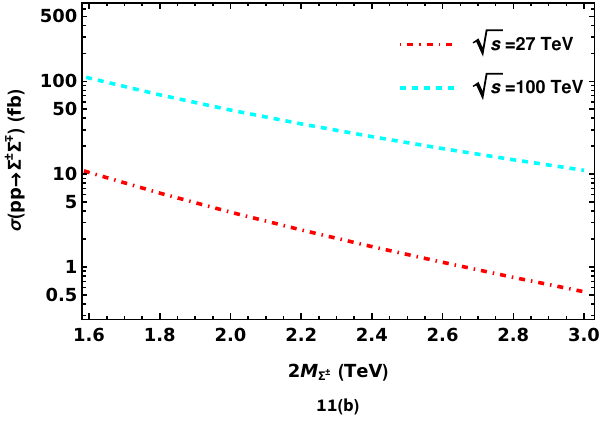}&\includegraphics[width=0.45\linewidth]{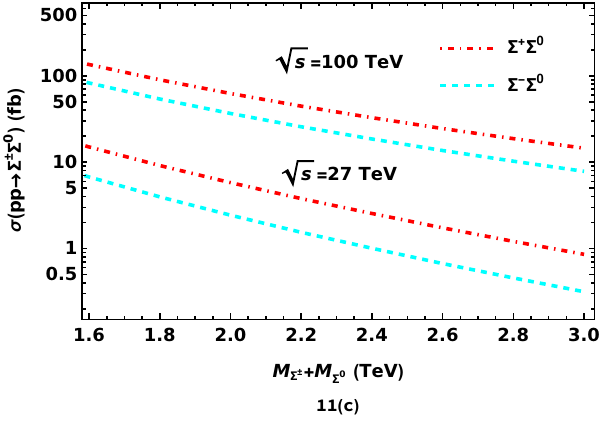}
	\end{tabular}
    
	\caption{Upper panel: pair-production cross-section for $\Sigma^\pm \Sigma^0$ and $\Sigma^{\pm} \Sigma^{\mp}$  as a function of   center-of-mass energy $(\sqrt{s})$   in $pp$ collisions. Lower panel: pair-production cross-section as a function of triplet fermion pair mass at HL-LHC and FCC-hh center-of-mass energy $(\sqrt{s})$, 27 TeV and 100 TeV.}
	\label{figtri_chafx2}
\end{figure}

The production modes for triplet fermions at $pp$ colliders are shown in Fig. \ref{fig_pp_tri_pro}.  Here, we can have pair-production of $\Sigma^\pm \Sigma^\mp$ ($\Sigma^\pm \Sigma^0$) via $s$-channel exchange of $Z$ boson and photon ($\gamma$) ($W^\pm$ boson). The production cross-section   in $pp$ colliders  as a function of $\sqrt{s}$ is shown in  Fig. \ref{figtri_chafx2}(a).  
The cross-section is found to be very small for low $\sqrt{s}$ and increases with $\sqrt{s}$. The cross-section of all three pairs is close to each other with $\Sigma^+ \Sigma^0$ dominating with value around $15$ fb at $\sqrt{s}=100$ TeV. In  Fig. \ref{figtri_chafx2}, we have shown the  production cross-section of fermion triplet pairs $\Sigma^\pm \Sigma^\mp$ (Fig. \ref{figtri_chafx2}(b)) and $\Sigma^\pm \Sigma^0$ (Fig. \ref{figtri_chafx2}(c)) for the HL-LHC and FCC-hh center-of-mass energies, $\sqrt{s}=27$ TeV and $100$ TeV. \\
We, further, study the signatures that can be seen at colliders from triplet fermion decays. The different decay chains for fermion triplet $\Sigma^\pm$ and $\Sigma^0$ along with their signatures considering inert scalar $H_1^0$ as DM are shown in Table \ref{tab9} and \ref{tab9b}, respectively.  The DM escapes without detection and appears  as missing transverse energy ($\cancel{E_T}$). By, combining these chains we can have  signatures in colliders which are tabulated, for $\Sigma^\pm \Sigma^\mp$ and  $\Sigma^\pm \Sigma^0$ in Table \ref{tab10} and \ref{tab10b}, respectively. In these tables the same colors of different cells shows, which combinations will give us the same signatures at colliders. For the case where $\Sigma^0$ is dark matter, $\Sigma^\pm$ can decay into $\Sigma^0$ and a $W^\pm$ boson. The $\Sigma^0$ appears as missing transverse energy ($\cancel{E_T}$). The $W^\pm$ further decays into either a lepton-neutrino pair ($l^\pm \nu_l$) or a quark-antiquark pair ($q \overline{q^\prime}$), leading to the signatures $1l + \cancel{E_T}$ and $2j + \cancel{E_T}$, respectively.

\begin{table}[h]
	\centering
	\begin{tabular}{|c|c|c|} \hline 
    Mode No. & Decay mode of $\Sigma^\pm$ & Signature\\ \hline
		1 & $H_1^0l^\pm$ & $1l+ \cancel{E_T}$\\
        2 & $(H_2^0/A_1^0/A_2^0) l^\pm \rightarrow H_1^0 h l^\pm \rightarrow H_1^0 q \overline{q^\prime}l^\pm$ & $1l+2j+ \cancel{E_T}$\\
        3 & $(H_2^0/A_1^0/A_2^0) l^\pm \rightarrow H_1^0 h l^\pm \rightarrow H_1^0 l^+ l^{\prime-} l^\pm$& $3l+\cancel{E_T}$\\
        4 & $(A_1^0/A_2^0) l^\pm \rightarrow H_1^0 Z l^\pm \rightarrow H_1^0 \nu_l \nu_l l^\pm$& $1l+\cancel{E_T}$\\
        5 & $(A_1^0/A_2^0) l^\pm \rightarrow H_1^0 Z l^\pm \rightarrow H_1^0 q \bar{q} l^\pm$& $1l+2j+\cancel{E_T}$\\
        6 & $(A_1^0/A_2^0) l^\pm \rightarrow H_1^0 Z l^\pm \rightarrow H_1^0 l^+ l^- l^\pm$& $3l+\cancel{E_T}$\\
        7 & $(H_1^\pm/H_2^\pm) \nu_l \rightarrow H_1^0 W^\pm \nu_l \rightarrow H_1^0l^\pm \nu_l \nu_l$& $1l+\cancel{E_T}$\\
        8 & $(H_1^\pm/H_2^\pm) \nu_l \rightarrow H_1^0 W^\pm \nu_l \rightarrow H_1^0 q \overline{q^\prime} \nu_l$& $2j+\cancel{E_T}$\\
        \hline
        \end{tabular}
	\caption{Decay modes of $\Sigma^\pm$ along with corresponding signature.}
	\label{tab9}
\end{table}

\begin{table}[h]
	\centering
	\begin{tabular}{|c|c|c|} \hline 
   Mode No. & Decay mode of $\Sigma^0$ & Signature\\ \hline
		$1^\prime$ & $H_1^0\nu_l$ & $\cancel{E_T}$\\
        $2^\prime$ & $(H_2^0/A_1^0/A_2^0) \nu_l \rightarrow H_1^0 h \nu_l \rightarrow H_1^0 q \overline{q^\prime}\nu_l$ & $2j+ \cancel{E_T}$\\
        $3^\prime$ & $(H_2^0/A_1^0/A_2^0) \nu_l \rightarrow H_1^0 h \nu_l \rightarrow H_1^0 l^+ l^{\prime-} \nu_l$& $2l+\cancel{E_T}$\\
        $4^\prime$ & $(A_1^0/A_2^0) \nu_l \rightarrow H_1^0 Z \nu_l \rightarrow H_1^0 \nu_l \nu_l \nu_l$& $\cancel{E_T}$\\
        $5^\prime$ & $(A_1^0/A_2^0) \nu_l \rightarrow H_1^0 Z \nu_l \rightarrow H_1^0 q \bar{q} \nu_l$& $2j+\cancel{E_T}$\\
        $6^\prime$ & $(A_1^0/A_2^0) \nu_l \rightarrow H_1^0 Z \nu_l \rightarrow H_1^0 l^+ l^- \nu_l$& $2l+\cancel{E_T}$\\
       $7^\prime$ & $(H_1^\pm/H_2^\pm) l^\mp \rightarrow H_1^0 W^\pm l^\mp \rightarrow H_1^0l^\pm \nu_l l^\mp$& $2l+\cancel{E_T}$\\
        $8^\prime$ & $(H_1^\pm/H_2^\pm) l^\mp \rightarrow H_1^0 W^\pm l^\mp \rightarrow H_1^0 q \overline{q^\prime} l^\mp$& $1l+2j+\cancel{E_T}$\\
        \hline
        \end{tabular}
	\caption{Decay modes of $\Sigma^0$ along with corresponding signature.}
	\label{tab9b}
\end{table}

\begin{table}[h]
\begin{scriptsize}
\centering
\begin{tabular}{|c|c|c|c|c|c|c|c|c|} \hline 
    $\frac{\Sigma^\mp\rightarrow}{\Sigma^\pm\downarrow}$&1 & 2 & 3 & 4 & 5 & 6 & 7 & 8\\ \hline
	1 & \cellcolor{cyan!30}$2l+\cancel{E_t}$ &\cellcolor{yellow!30}$2l+2j+\cancel{E_t}$ &\cellcolor{green!30}$4l+\cancel{E_t}$ & \cellcolor{cyan!30}$2l+\cancel{E_t}$ &\cellcolor{yellow!30}$2l+ 2j + \cancel{E_t}$ & \cellcolor{green!30}$4l+\cancel{E_t}$ & \cellcolor{cyan!30}$2l+\cancel{E_t}$ & \cellcolor{red!30}$1l+2j+\cancel{E_t}$\\
    2 & \cellcolor{yellow!30}$2l+2j+\cancel{E_t}$ & \cellcolor{blue!30}$2l+4j+ \cancel{E_T}$& \cellcolor{orange!30}$4l+2j+ \cancel{E_T}$ & \cellcolor{yellow!30}$2l+2j+ \cancel{E_T}$ & \cellcolor{blue!30}$2l+4j+ \cancel{E_T}$ & \cellcolor{orange!30}$4l+2j+ \cancel{E_T}$ & \cellcolor{yellow!30}$2l+2j+ \cancel{E_T}$ & \cellcolor{gray!25}$1l+4j+ \cancel{E_T}$\\
    3 & \cellcolor{green!30}$4l+\cancel{E_t}$ & \cellcolor{orange!30}$4l+ 2j+\cancel{E_T}$ & \cellcolor{pink!30}$6l+\cancel{E_T}$ & \cellcolor{green!30}$4l+\cancel{E_T}$ & \cellcolor{orange!30}$4l+ 2j+\cancel{E_T}$ & \cellcolor{pink!30}$6l+\cancel{E_T}$  & \cellcolor{green!30}$4l+\cancel{E_T}$ & \cellcolor{black!30}$3l+ 2j+\cancel{E_T}$   \\
    4 & \cellcolor{cyan!30}$2l+\cancel{E_t}$& \cellcolor{yellow!30}$2l+2j+\cancel{E_T}$ & \cellcolor{green!30}$4l+\cancel{E_T}$ & \cellcolor{cyan!30}$2l+\cancel{E_T}$ & \cellcolor{yellow!30}$3l+ 2j+\cancel{E_T}$ & \cellcolor{green!30}$4l+\cancel{E_T}$ & \cellcolor{cyan!30}$2l+\cancel{E_T}$ & \cellcolor{red!30}$1l+ 2j+\cancel{E_T}$    \\
    5 & \cellcolor{yellow!30}$2l+2j+\cancel{E_t}$& \cellcolor{blue!30}$2l+4j+\cancel{E_T}$ & \cellcolor{orange!30}$4l+ 2j+\cancel{E_T}$ & \cellcolor{yellow!30}$2l+ 2j+\cancel{E_T}$  & \cellcolor{blue!30}$2l+ 4j+\cancel{E_T}$ & \cellcolor{orange!30}$4l+ 2j+\cancel{E_T}$ & \cellcolor{yellow!30}$2l+ 2j+\cancel{E_T}$ & \cellcolor{gray!25}$1l+ 4j+\cancel{E_T}$   \\
    6 & \cellcolor{green!30}$4l+\cancel{E_t}$& \cellcolor{orange!30}$4l+2j+\cancel{E_T}$ & \cellcolor{pink!30}$6l+\cancel{E_T}$ & \cellcolor{green!30}$4l+ \cancel{E_T}$ & \cellcolor{orange!30}$4l+ 2j+\cancel{E_T}$ & \cellcolor{pink!30}$6l+\cancel{E_T}$  & \cellcolor{green!30}$4l+\cancel{E_T}$  & \cellcolor{black!30}$3l+ 2j+\cancel{E_T}$  \\
    7 & \cellcolor{cyan!30}$2l+\cancel{E_t}$& \cellcolor{yellow!30}$2l+2j+\cancel{E_T}$ & \cellcolor{green!30}$4l+\cancel{E_T}$ & \cellcolor{cyan!30}$2l+ \cancel{E_T}$ & \cellcolor{yellow!30}$2l+ 2j+\cancel{E_T}$ & \cellcolor{green!30}$4l+\cancel{E_T}$  & \cellcolor{cyan!30}$2l+\cancel{E_T}$  & \cellcolor{red!30}$1l+ 2j+\cancel{E_T}$  \\
    8 & \cellcolor{red!30}$1l+2j+\cancel{E_t}$& \cellcolor{gray!25}$1l+4j+\cancel{E_T}$ & \cellcolor{black!30}$3l+ 2j+\cancel{E_T}$ & \cellcolor{red!30}$1l+ 2j+\cancel{E_T}$  & \cellcolor{gray!25}$1l+ 4j+\cancel{E_T}$ & \cellcolor{black!30}$3l+ 2j+\cancel{E_T}$ & \cellcolor{red!30}$1l+ 2j+\cancel{E_T}$ & \cellcolor{brown!60}$ 4j+\cancel{E_T}$   \\
    \hline
    \end{tabular}
\caption{Possible signatures of $\Sigma^\pm \Sigma^\mp$ pair by combining $\Sigma^\pm$ (1st column) and $\Sigma^\mp$ (1st row) signatures for the  decays shown in Table \ref{tab9}.}
\label{tab10}
\end{scriptsize}   
\end{table}
\FloatBarrier
\begin{table}[h]
\begin{scriptsize}
\centering
\begin{tabular}{|c|c|c|c|c|c|c|c|c|} \hline 
    $\frac{\Sigma^0\rightarrow}{\Sigma^\pm\downarrow}$&$1^\prime$ &$2^\prime$ & $3^\prime$ & $4^\prime$ & $5^\prime$ & $6^\prime$ & $7^\prime$ & $8^\prime$\\ \hline
	1 & \cellcolor{cyan!30}$1l+\cancel{E_t}$& \cellcolor{yellow!30}$1l+2j+\cancel{E_t}$& \cellcolor{green!30}$3l+\cancel{E_t}$& \cellcolor{cyan!30}$1l+\cancel{E_t}$& \cellcolor{yellow!30}$1l+ 2j + \cancel{E_t}$& \cellcolor{green!30}$3l+\cancel{E_t}$& \cellcolor{green!30}$3l+\cancel{E_t}$& \cellcolor{red!30}$2l+2j+\cancel{E_t}$\\
    2 & \cellcolor{yellow!30}$1l+2j+\cancel{E_t}$ & \cellcolor{blue!30}$1l+4j+ \cancel{E_T}$& \cellcolor{orange!30}$3l+2j+ \cancel{E_T}$& \cellcolor{yellow!30}$1l+2j+ \cancel{E_T}$& \cellcolor{blue!30}$1l+4j+ \cancel{E_T}$& \cellcolor{orange!30}$3l+2j+ \cancel{E_T}$& \cellcolor{orange!30}$3l+2j+ \cancel{E_T}$& \cellcolor{gray!25}$2l+4j+ \cancel{E_T}$\\
    3 & \cellcolor{green!30}$3l+\cancel{E_t}$ & \cellcolor{orange!30}$3l+2j+\cancel{E_T}$ & \cellcolor{pink!30}$5l+\cancel{E_T}$& \cellcolor{green!30}$3l+\cancel{E_T}$& \cellcolor{orange!30}$3l+ 2j+\cancel{E_T}$& \cellcolor{pink!30}$5l+\cancel{E_T}$& \cellcolor{pink!30}$5l+\cancel{E_T}$& \cellcolor{black!30}$4l+ 2j+\cancel{E_T}$\\
    4 & \cellcolor{cyan!30}$1l+\cancel{E_t}$& \cellcolor{yellow!30}$1l+2j+\cancel{E_T}$ & \cellcolor{green!30}$3l+\cancel{E_T}$ & \cellcolor{cyan!30}$1l+\cancel{E_T}$& \cellcolor{yellow!30}$1l+ 2j+\cancel{E_T}$& \cellcolor{green!30}$3l+\cancel{E_T}$& \cellcolor{green!30}$3l+\cancel{E_T}$& \cellcolor{red!30}$2l+ 2j+\cancel{E_T}$\\
    5 & \cellcolor{yellow!30}$1l+2j+\cancel{E_t}$& \cellcolor{blue!30}$1l+4j+\cancel{E_T}$ & \cellcolor{orange!30}$3l+ 2j+\cancel{E_T}$ & \cellcolor{yellow!30}$1l+ 2j+\cancel{E_T}$  & \cellcolor{blue!30}$1l+ 4j+\cancel{E_T}$& \cellcolor{orange!30}$3l+ 2j+\cancel{E_T}$& \cellcolor{orange!30}$3l+ 2j+\cancel{E_T}$& \cellcolor{gray!25}$2l+ 4j+\cancel{E_T}$\\
    6 & \cellcolor{green!30}$3l+\cancel{E_t}$& \cellcolor{orange!30}$3l+2j+\cancel{E_T}$ & \cellcolor{pink!30}$5l+\cancel{E_T}$ & \cellcolor{green!30}$3l+ \cancel{E_T}$ & \cellcolor{orange!30}$3l+ 2j+\cancel{E_T}$ & \cellcolor{pink!30}$5l+\cancel{E_T}$& \cellcolor{pink!30}$5l+\cancel{E_T}$& \cellcolor{black!30}$4l+ 2j+\cancel{E_T}$  \\
    7 & \cellcolor{green!30}$3l+\cancel{E_t}$& \cellcolor{orange!30}$3l+2j+\cancel{E_T}$ & \cellcolor{pink!30}$5l+\cancel{E_T}$ & \cellcolor{green!30}$3l+ \cancel{E_T}$ & \cellcolor{orange!30}$3l+ 2j+\cancel{E_T}$ & \cellcolor{pink!30}$5l+\cancel{E_T}$  & \cellcolor{green!30}$3l+\cancel{E_T}$& \cellcolor{red!30}$2l+ 2j+\cancel{E_T}$\\
    8 & \cellcolor{red!30}$2l+2j+\cancel{E_t}$& \cellcolor{gray!25}$2l+4j+\cancel{E_T}$ & \cellcolor{black!30}$4l+ 2j+\cancel{E_T}$ & \cellcolor{red!30}$2l+ 2j+\cancel{E_T}$  & \cellcolor{gray!25}$2l+ 4j+\cancel{E_T}$ & \cellcolor{black!30}$4l+ 2j+\cancel{E_T}$ & \cellcolor{red!30}$2l+ 2j+\cancel{E_T}$ & \cellcolor{blue!30}$1l+4j+\cancel{E_T}$\\
    \hline
    \end{tabular}
\caption{Possible signatures for  $\Sigma^\pm \Sigma^0$ pair by combining $\Sigma^\pm$ (1st column) and $\Sigma^0$ (1st row) signatures for the  decays shown in Table \ref{tab9} and Table \ref{tab9b}, respectively.}
\label{tab10b}
\end{scriptsize}   
\end{table}

\section{Conclusions}
\noindent In this work, we explore an extended radiative Type-III scotogenic model featuring two inert scalar doublets and a fermion triplet, where neutrino masses arise at the one-loop level with both inert doublets circulating in the loop. The scalar sector, in addition to the SM Higgs, consists of a rich spectrum of dark scalars: two CP-even, two CP-odd, and two charged states. This framework accommodates two viable DM candidates: the CP-even inert scalar $H_1^0$ and the neutral component of the triplet fermion $\Sigma^0$. A comprehensive analysis of dark matter phenomenology reveals new viable mass regions where the observed DM relic density is satisfied for both scenarios, shaped by intricate co-annihilation dynamics. In the inert scalar DM scenario, a significant parameter space emerges below 500 GeV—a region often dubbed the ``desert region'' in the conventional inert doublet model. In the fermionic DM scenario, a new viable region is identified for DM mass below 2.5 TeV. These regions materialize under specific mass hierarchies, where all inert scalars remain nearly degenerate with the DM candidate (Case III for inert scalar DM and Case II for triplet fermion DM in Table \ref{table2}).

\noindent We have also explored the collider implications of the model at both $pp$ and $e^{+}e^{-}$ colliders, focusing on the missing energy signatures of the triplet fermion. At higher mass scale, the production cross-section of triplet fermion drops significantly, making detection challenging at the LHC with $\sqrt{s}=13$ TeV. However, at a future 100 TeV proton-proton collider like FCC-hh, the production cross-section remains substantially high, offering a promising avenue for discovery. This enhancement, as shown in the upper panel of Fig. \ref{figtri_chafx2}, underscores the FCC-hh’s potential to probe the triplet fermion. Further, we also explored the possible signatures of the triplet fermion at $e^{+}e^{-}$ collider, such as ILC and CLIC. Tables \ref{tab10} and \ref{tab10b} comprise the possible signatures of triplet fermion at both $pp$ and $e^{+}e^{-}$ colliders.

\noindent\textbf{\Large{Acknowledgments}}
 \vspace{.3cm}\\
Tapender acknowledges the financial support provided by Central University of Himachal Pradesh in the form of freeship. LS acknowledges the financial support provided by the Council of Scientific and Industrial Research (CSIR) vide letter No. 09/1196(18553)/2024-EMR-I. The authors, also, acknowledge Department of Physics and Astronomical Science for providing necessary facility to carry out this work.

\begin{appendices}
\section{Feynman diagrams for various processes}\label{appendix}
\begin{figure}[hbt!]
	\centering
	\begin{tikzpicture}[scale=0.6]
		\begin{feynman}
			\vertex at (0,0) (i1);
			\vertex at (-1.5,0) (i2);
			\vertex at (1.5,1.5) (a);
			\vertex at (1.5,-1.5) (b);
			\vertex at (-2.5,1.5) (c);
			\vertex at (-2.5,-1.5) (d);
			
			\vertex at (-2.7,-1.7) () {\(S\)};
			\vertex at (-2.7,1.7) () {\(S\)};
			\vertex at (1.7,1.7) () {\(f\)};
			\vertex at (1.7,-1.7) () {\( \bar{f}, \bar{f'} \)};
			\vertex at (-0.8,-0.2) () {\(h\)};
			\diagram*{
				(i2) -- [scalar] (i1), (i1) -- [fermion] (a), (b) -- [fermion] (i1), (i2) -- [scalar] (c),(d) -- [scalar] (i2)
			};
		\end{feynman}
		\node at (-0.6, -2.5) {(a)};
	\end{tikzpicture}
	\hspace{0.3cm}
	\begin{tikzpicture}[scale=0.6]
		\begin{feynman}
			\vertex at (0,0) (i1);
			\vertex at (-1.5,0) (i2);
			\vertex at (1.5,1.5) (a);
			\vertex at (1.5,-1.5) (b);
			\vertex at (-2.5,1.5) (c);
			\vertex at (-2.5,-1.5) (d);
			
			\vertex at (-2.7,-1.7) () {\(S\)};
			\vertex at (-2.7,1.7) () {\(S\)};
			\vertex at (1.7,1.7) () {\(g\)};
			\vertex at (1.7,-1.7) () {\(g \)};
			\vertex at (-0.5,-0.2) () {\(h\)};
			\diagram*{
				(i2) -- [scalar] (i1), (i1) -- [gluon] (a), (b) -- [gluon] (i1), (i2) -- [scalar] (c),(d) -- [scalar] (i2)
			};
		\end{feynman}
		\node at (-0.6, -2.5) {(b)};
	\end{tikzpicture}
	\hspace{0.3cm}
	\begin{tikzpicture}[scale=0.6]
		\begin{feynman}
			\vertex at (0,0) (a);
			\vertex at (1.7,-1) (i1);
			\vertex at (-1.7,-1) (i2);
			\vertex at (0,1.5) (b);
			\vertex at (1.7,2.5) (c);
			\vertex at (-1.7,2.5) (d);
			\vertex at (1.8,2.7) () {\(l^+(l^-,l'^-)\)};
			\vertex at (-1.8,2.7) () {\(S\)};
			\vertex at (1.8,-1.3) () {\(l^-,l'^-(l^+)\)};
			\vertex at (-1.8,-1.3) () {\(S\)};
			\vertex at (1.3,1) () {\( \Sigma^+(\Sigma^-)\)};
			
			\diagram*{
				(a) -- [] (i1), (i2) -- [scalar] (a),
				(b) -- [] (a), (c) -- [] (b), (d) -- [scalar] (b),
			};
		\end{feynman}
		\node at (0, -2.5) {(c)};
	\end{tikzpicture}
	
	\begin{tikzpicture}[scale=0.6]
		\begin{feynman}
			\vertex at (0,0) (i1);
			\vertex at (2.2,1.7) (a);
			\vertex at (2.2,-1.5) (b);
			\vertex at (-2.2,1.7) (c);
			\vertex at (-2.2,-1.5) (d);
			
			\vertex at (-2.4,-1.7) () {\(S\)};
			\vertex at (-2.4,1.9) () {\(S\)};
			\vertex at (2.4,1.9) () {\(Z (W^+)\)};
			\vertex at (2.4,-1.7) () {\(Z (W^-)\)};
			\diagram*{
				(i1) -- [photon] (a), (b) -- [photon] (i1), (c) -- [scalar] (i1),(d) -- [scalar] (i1)
			};
		\end{feynman}
		\node at (0, -2.5) {(d)};
	\end{tikzpicture}
	\hspace{0.3cm}
	\begin{tikzpicture}[scale=0.6]
		\begin{feynman}
			\vertex at (0,0) (i1);
			\vertex at (-1.5,0) (i2);
			\vertex at (1.5,1.5) (a);
			\vertex at (1.5,-1.5) (b);
			\vertex at (-2.5,1.5) (c);
			\vertex at (-2.5,-1.5) (d);
			
			\vertex at (-2.7,-1.7) () {\(S\)};
			\vertex at (-2.7,1.7) () {\(S\)};
			\vertex at (1.7,1.7) () {\(Z (W^+)\)};
			\vertex at (1.7,-1.7) () {\( Z(W^-) \)};
			\vertex at (-0.5,-0.2) () {\(h\)};
			\diagram*{
				(i2) -- [scalar] (i1), (i1) -- [photon] (a), (b) -- [photon] (i1), (i2) -- [scalar] (c),(d) -- [scalar] (i2)
			};
		\end{feynman}
		\node at (-0.6, -2.5) {(e)};
	\end{tikzpicture}
	\hspace{0.3cm}
	\begin{tikzpicture}[scale=0.6]
		\begin{feynman}
			\vertex at (0,0) (a);
			\vertex at (1.7,-1) (i1);
			\vertex at (-1.7,-1) (i2);
			\vertex at (0,1.5) (b);
			\vertex at (1.7,2.5) (c);
			\vertex at (-1.7,2.5) (d);
			\vertex at (1.8,2.7) () {\(Z(W^\pm)\)};
			\vertex at (-1.8,2.7) () {\(S\)};
			\vertex at (1.8,-1.3) () {\(Z(W^\mp)\)};
			\vertex at (-1.8,-1.3) () {\(S\)};
			\vertex at (1.5,0.7) () {\(A_{1,2}^0(H_{1,2}^\pm) \)};
			
			\diagram*{
				(a) -- [photon] (i1), (i2) -- [scalar] (a),
				(b) -- [scalar] (a), (c) -- [photon] (b), (d) -- [scalar] (b),
			};
		\end{feynman}
		\node at (0, -2.5) {(f)};
	\end{tikzpicture}	
	\centering
	\begin{tikzpicture}[scale=0.6]
		\begin{feynman}
			\vertex at (0,0) (i1);
			\vertex at (-1.5,0) (i2);
			\vertex at (1.5,1.5) (a);
			\vertex at (1.5,-1.5) (b);
			\vertex at (-2.5,1.5) (c);
			\vertex at (-2.5,-1.5) (d);
			
			\vertex at (-2.7,-1.7) () {\(S\)};
			\vertex at (-2.7,1.7) () {\(S\)};
			\vertex at (1.7,1.7) () {\(h\)};
			\vertex at (1.7,-1.7) () {\( h \)};
			\vertex at (-0.8,-0.2) () {\(h\)};
			\diagram*{
				(i2) -- [scalar] (i1), (i1) -- [scalar] (a), (b) -- [scalar] (i1), (i2) -- [scalar] (c),(d) -- [scalar] (i2)
			};
		\end{feynman}
		\node at (-0.6, -2.5) {(g)};
	\end{tikzpicture}
	\hspace{0.3cm}
	\begin{tikzpicture}[scale=0.6]
		\begin{feynman}
			\vertex at (0,0) (a);
			\vertex at (1.7,-1) (i1);
			\vertex at (-1.7,-1) (i2);
			\vertex at (0,1.5) (b);
			\vertex at (1.7,2.5) (c);
			\vertex at (-1.7,2.5) (d);
			\vertex at (1.8,2.7) () {\(h\)};
			\vertex at (-1.8,2.7) () {\(S\)};
			\vertex at (1.8,-1.3) () {\(h\)};
			\vertex at (-1.8,-1.3) () {\(S\)};
			\vertex at (1.3,1) () {\( H_{1,2}^0, A_{1,2}^0 \)};
			
			\diagram*{
				(a) -- [scalar] (i1), (i2) -- [scalar] (a),
				(b) -- [scalar] (a), (c) -- [scalar] (b), (d) -- [scalar] (b),
			};
		\end{feynman}
		\node at (0, -2.5) {(h)};
	\end{tikzpicture}
		%
	\caption{Dominating annihilation channels where $S$ can be $H^0_{1,2},A^0_{1,2}$, where $f$ ($l$) represent quark (lepton).}
	\label{fig6}
\end{figure}
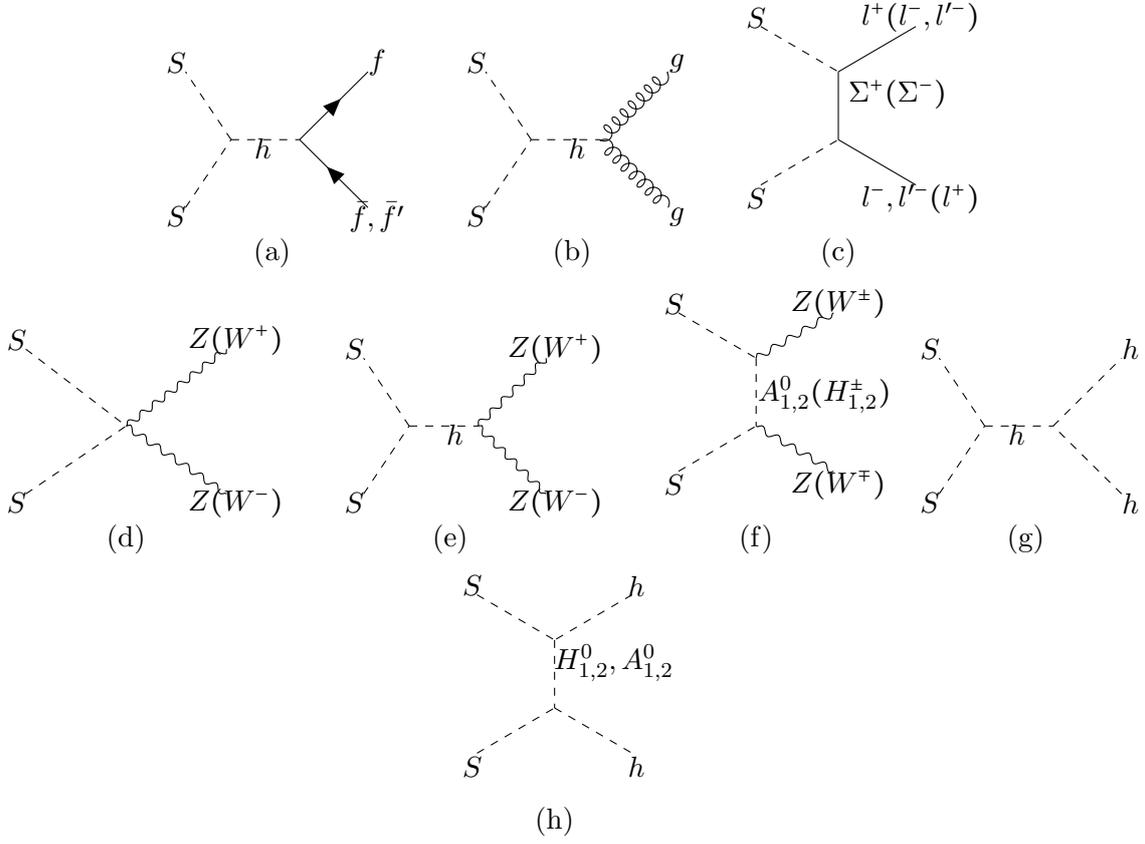


\begin{figure}[hbt!]
	\centering
	\begin{tikzpicture}[scale=0.6]
		\begin{feynman}
			\vertex at (0,0) (i1);
			\vertex at (-1.5,0) (i2);
			\vertex at (1.5,1.5) (a);
			\vertex at (1.5,-1.5) (b);
			\vertex at (-2.5,1.5) (c);
			\vertex at (-2.5,-1.5) (d);
			
			\vertex at (-2.7,-1.7) () {\(H^0_1\)};
			\vertex at (-2.7,1.7) () {\(A^0_1\)};
			\vertex at (1.7,1.7) () {\(Z\)};
			\vertex at (1.7,-1.7) () {\( h \)};
			\vertex at (-0.8,-0.2) () {\(Z\)};
			\diagram*{
				(i2) -- [scalar] (i1), (i1) -- [scalar] (a), (b) -- [scalar] (i1), (i2) -- [scalar] (c),(d) -- [scalar] (i2)
			};
		\end{feynman}
        \node at (0, -2.5) {(a)};
	\end{tikzpicture}
	\hspace{0.3cm}
	\begin{tikzpicture}[scale=0.6]
		\begin{feynman}
			\vertex at (0,0) (a);
			\vertex at (1.7,-1) (i1);
			\vertex at (-1.7,-1) (i2);
			\vertex at (0,1.5) (b);
			\vertex at (1.7,2.5) (c);
			\vertex at (-1.7,2.5) (d);
			\vertex at (1.8,2.7) () {\(Z\)};
			\vertex at (-1.8,2.7) () {\(A^0_1\)};
			\vertex at (1.8,-1.3) () {\(h\)};
			\vertex at (-1.8,-1.3) () {\(H^0_1\)};
			\vertex at (0.8,1) () {\( H_{1,2}^0 \)};
			
			\diagram*{
				(a) -- [scalar] (i1), (i2) -- [scalar] (a),
				(b) -- [scalar] (a), (c) -- [scalar] (b), (d) -- [scalar] (b),
			};
		\end{feynman}
        \node at (0, -2.5) {(b)};
	\end{tikzpicture}
	\hspace{0.3cm}
	\begin{tikzpicture}[scale=0.6]
		\begin{feynman}
			\vertex at (0,0) (a);
			\vertex at (1.7,-1) (i1);
			\vertex at (-1.7,-1) (i2);
			\vertex at (0,1.5) (b);
			\vertex at (1.7,2.5) (c);
			\vertex at (-1.7,2.5) (d);
			\vertex at (1.8,2.7) () {\(h\)};
			\vertex at (-1.8,2.7) () {\(A^0_1\)};
			\vertex at (1.8,-1.3) () {\(Z\)};
			\vertex at (-1.8,-1.3) () {\(H^0_1\)};
			\vertex at (0.8,1) () {\( A_{1,2}^0 \)};
			
			\diagram*{
				(a) -- [scalar] (i1), (i2) -- [scalar] (a),
				(b) -- [scalar] (a), (c) -- [scalar] (b), (d) -- [scalar] (b),
			};
		\end{feynman}
        \node at (0, -2.5) {(c)};
	\end{tikzpicture}
    	\centering
	\begin{tikzpicture}[scale=0.6]
		\begin{feynman}
			\vertex at (0,0) (i1);
			\vertex at (-1.5,0) (i2);
			\vertex at (1.5,1.5) (a);
			\vertex at (1.5,-1.5) (b);
			\vertex at (-2.5,1.5) (c);
			\vertex at (-2.5,-1.5) (d);
			
			\vertex at (-2.7,-1.7) () {\(S\)};
			\vertex at (-2.7,1.7) () {\(H^\pm_{1,2}\)};
			\vertex at (1.7,1.9) () {\(W^+\)};
			\vertex at (1.7,-1.7) () {\( h \)};
			\vertex at (-0.8,-0.6) () {\(W^+\)};
			\diagram*{
				(i2) -- [photon] (i1), (i1) -- [photon] (a), (b) -- [scalar] (i1), (i2) -- [scalar] (c),(d) -- [scalar] (i2)
			};
		\end{feynman}
        \node at (0, -2.5) {(d)};
	\end{tikzpicture}
	\hspace{0.3cm}
	\begin{tikzpicture}[scale=0.6]
		\begin{feynman}
			\vertex at (0,0) (a);
			\vertex at (1.7,-1) (i1);
			\vertex at (-1.7,-1) (i2);
			\vertex at (0,1.5) (b);
			\vertex at (1.7,2.5) (c);
			\vertex at (-1.7,2.5) (d);
			\vertex at (1.8,2.7) () {\(W^+\)};
			\vertex at (-1.8,2.7) () {\(H^\pm_{1,2} \)};
			\vertex at (1.8,-1.3) () {\(h\)};
			\vertex at (-1.8,-1.3) () {\(S\)};
			\vertex at (1.5,1) () {\( H_{1,2}^0,A^0_{1,2} \)};
			
			\diagram*{
				(a) -- [scalar] (i1), (i2) -- [scalar] (a),
				(b) -- [scalar] (a), (c) -- [scalar] (b), (d) -- [scalar] (b),
			};
		\end{feynman}
        \node at (0, -2.5) {(e)};
	\end{tikzpicture}
	\hspace{0.3cm}
	\begin{tikzpicture}[scale=0.6]
		\begin{feynman}
			\vertex at (0,0) (a);
			\vertex at (1.7,-1) (i1);
			\vertex at (-1.7,-1) (i2);
			\vertex at (0,1.5) (b);
			\vertex at (1.7,2.5) (c);
			\vertex at (-1.7,2.5) (d);
			\vertex at (1.8,2.7) () {\(h\)};
			\vertex at (-1.8,2.7) () {\(H^\pm_{1,2}\)};
			\vertex at (1.8,-1.3) () {\(W^+\)};
			\vertex at (-1.8,-1.3) () {\(S\)};
			\vertex at (0.8,1) () {\( H_{1,2}^{\mp}\)};
			
			\diagram*{
				(a) -- [scalar] (i1), (i2) -- [scalar] (a),
				(b) -- [scalar] (a), (c) -- [scalar] (b), (d) -- [scalar] (b),
			};
		\end{feynman}
        \node at (0, -2.5) {(f)};
	\end{tikzpicture}
	\caption{Dominating co-annihilation channels where $S$ can be $H^0_{1,2},A^0_{1,2}$.}
	\label{fig8}
\end{figure}
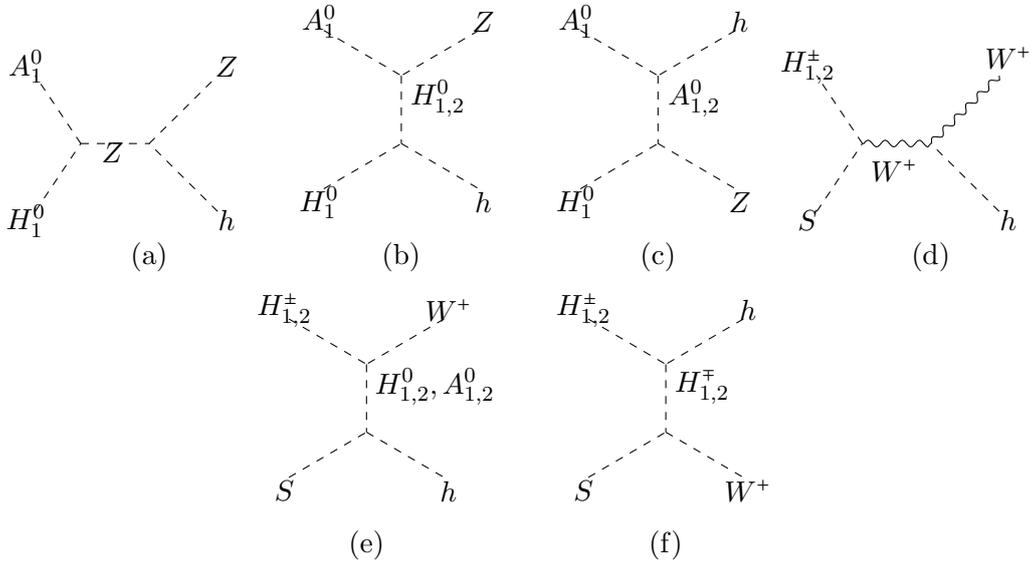


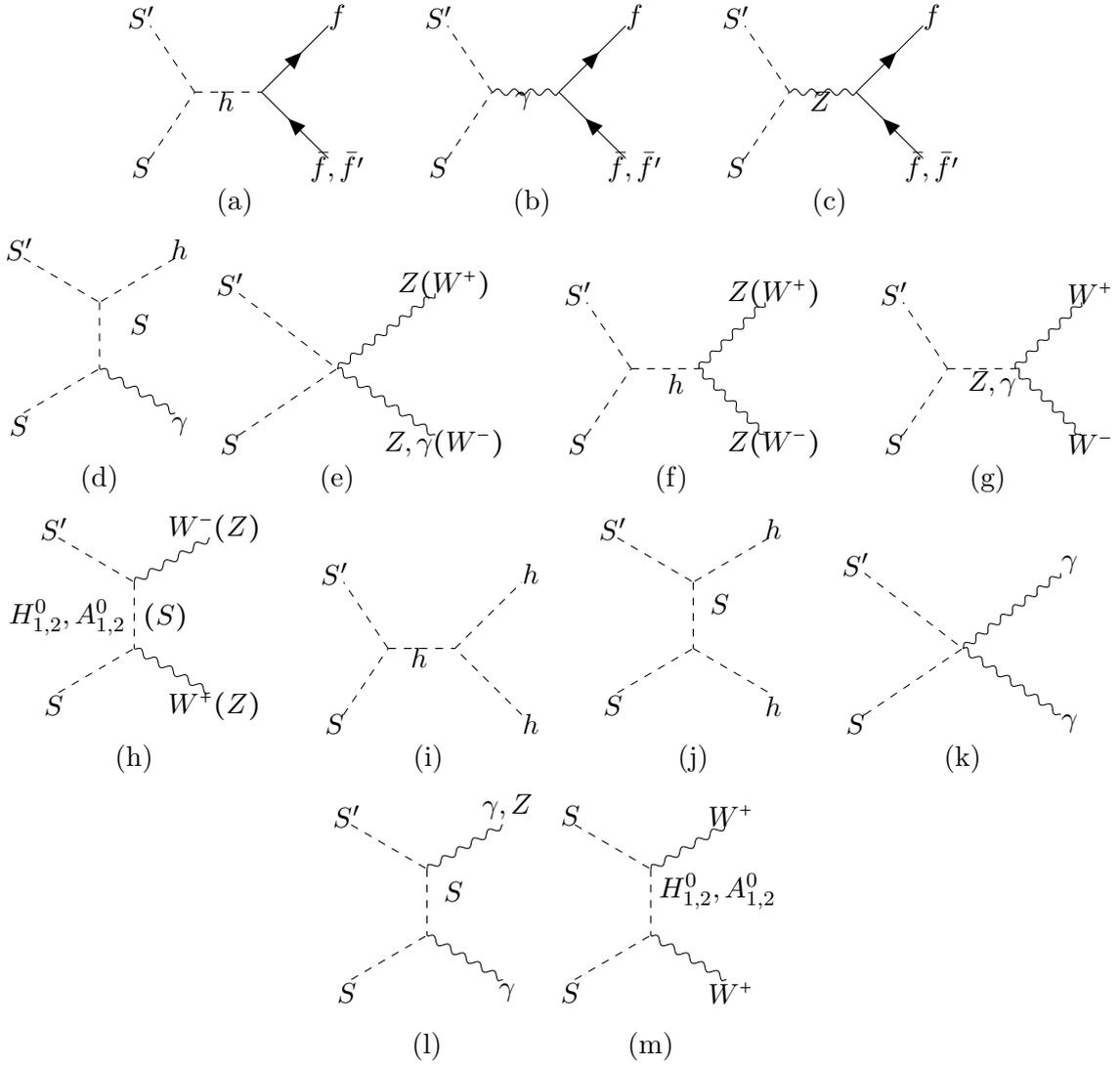
\begin{figure}[hbt!]
	\centering
	\begin{tikzpicture}[scale=0.6]
		\begin{feynman}
			\vertex at (0,0) (i1);
			\vertex at (-1.5,0) (i2);
			\vertex at (1.5,1.5) (a);
			\vertex at (1.5,-1.5) (b);
			\vertex at (-2.5,1.5) (c);
			\vertex at (-2.5,-1.5) (d);
			
			\vertex at (-2.7,-1.7) () {\(S\)};
			\vertex at (-2.7,1.7) () {\(S^\prime\)};
			\vertex at (1.7,1.7) () {\(f\)};
			\vertex at (1.7,-1.7) () {\( \bar{f}, \bar{f'} \)};
			\vertex at (-0.8,-0.2) () {\(h\)};
			\diagram*{
				(i2) -- [scalar] (i1), (i1) -- [fermion] (a), (b) -- [fermion] (i1), (i2) -- [scalar] (c),(d) -- [scalar] (i2)
			};
		\end{feynman}
		\node at (-0.6, -2.5) {(a)};
	\end{tikzpicture}
	\hspace{0.3cm}
	\begin{tikzpicture}[scale=0.6]
		\begin{feynman}
			\vertex at (0,0) (i1);
			\vertex at (-1.5,0) (i2);
			\vertex at (1.5,1.5) (a);
			\vertex at (1.5,-1.5) (b);
			\vertex at (-2.5,1.5) (c);
			\vertex at (-2.5,-1.5) (d);
			
			\vertex at (-2.7,-1.7) () {\(S\)};
			\vertex at (-2.7,1.7) () {\(S^\prime\)};
			\vertex at (1.7,1.7) () {\(f\)};
			\vertex at (1.7,-1.7) () {\( \bar{f}, \bar{f'} \)};
			\vertex at (-0.8,-0.2) () {\(\gamma\)};
			\diagram*{
				(i2) -- [boson] (i1), (i1) -- [fermion] (a), (b) -- [fermion] (i1), (i2) -- [scalar] (c),(d) -- [scalar] (i2)
			};
		\end{feynman}
		\node at (-0.6, -2.5) {(b)};
	\end{tikzpicture}
	\hspace{0.3cm}
	\begin{tikzpicture}[scale=0.6]
		\begin{feynman}
			\vertex at (0,0) (i1);
			\vertex at (-1.5,0) (i2);
			\vertex at (1.5,1.5) (a);
			\vertex at (1.5,-1.5) (b);
			\vertex at (-2.5,1.5) (c);
			\vertex at (-2.5,-1.5) (d);
			
			\vertex at (-2.7,-1.7) () {\(S\)};
			\vertex at (-2.7,1.7) () {\(S^\prime\)};
			\vertex at (1.7,1.7) () {\(f\)};
			\vertex at (1.7,-1.7) () {\( \bar{f}, \bar{f'} \)};
			\vertex at (-0.8,-0.2) () {\(Z\)};
			\diagram*{
				(i2) -- [boson] (i1), (i1) -- [fermion] (a), (b) -- [fermion] (i1), (i2) -- [scalar] (c),(d) -- [scalar] (i2)
			};
		\end{feynman}
		\node at (-0.6, -2.5) {(c)};
	\end{tikzpicture}
			
	\hspace{0.3cm}
	\begin{tikzpicture}[scale=0.6]
		\begin{feynman}
			\vertex at (0,0) (a);
			\vertex at (1.7,-1) (i1);
			\vertex at (-1.7,-1) (i2);
			\vertex at (0,1.5) (b);
			\vertex at (1.7,2.5) (c);
			\vertex at (-1.7,2.5) (d);
			\vertex at (1.8,2.7) () {\(h\)};
			\vertex at (-1.8,2.7) () {\(S^\prime\)};
			\vertex at (1.8,-1.3) () {\(\gamma\)};
			\vertex at (-1.8,-1.3) () {\(S\)};
			\vertex at (0.9,1) () {\( S \)};
			
			\diagram*{
				(a) -- [boson] (i1), (i2) -- [scalar] (a),
				(b) -- [scalar] (a), (c) -- [scalar] (b), (d) -- [scalar] (b),
			};
		\end{feynman}
		\node at (0, -2.5) {(d)};
	\end{tikzpicture}
	\begin{tikzpicture}[scale=0.6]
		\begin{feynman}
			\vertex at (0,0) (i1);
			\vertex at (2.2,1.7) (a);
			\vertex at (2.2,-1.5) (b);
			\vertex at (-2.2,1.7) (c);
			\vertex at (-2.2,-1.5) (d);
			
			\vertex at (-2.4,-1.7) () {\(S\)};
			\vertex at (-2.4,1.9) () {\(S^\prime \)};
			\vertex at (2.4,1.9) () {\(Z (W^+)\)};
			\vertex at (2.4,-1.7) () {\(Z,\gamma (W^-)\)};
			\diagram*{
				(i1) -- [photon] (a), (b) -- [photon] (i1), (c) -- [scalar] (i1),(d) -- [scalar] (i1)
			};
		\end{feynman}
		\node at (0, -2.5) {(e)};
	\end{tikzpicture}
	\hspace{0.3cm}
	\begin{tikzpicture}[scale=0.6]
		\begin{feynman}
			\vertex at (0,0) (i1);
			\vertex at (-1.5,0) (i2);
			\vertex at (1.5,1.5) (a);
			\vertex at (1.5,-1.5) (b);
			\vertex at (-2.5,1.5) (c);
			\vertex at (-2.5,-1.5) (d);
			
			\vertex at (-2.7,-1.7) () {\(S\)};
			\vertex at (-2.7,1.7) () {\(S^\prime\)};
			\vertex at (1.7,1.7) () {\(Z (W^+)\)};
			\vertex at (1.7,-1.7) () {\( Z(W^-) \)};
			\vertex at (-0.5,-0.35) () {\(h\)};
			\diagram*{
				(i2) -- [scalar] (i1), (i1) -- [photon] (a), (b) -- [photon] (i1), (i2) -- [scalar] (c),(d) -- [scalar] (i2)
			};
		\end{feynman}
		\node at (-0.6, -2.5) {(f)};
	\end{tikzpicture}
	\hspace{0.3cm}
	\begin{tikzpicture}[scale=0.6]
		\begin{feynman}
			\vertex at (0,0) (i1);
			\vertex at (-1.5,0) (i2);
			\vertex at (1.5,1.5) (a);
			\vertex at (1.5,-1.5) (b);
			\vertex at (-2.5,1.5) (c);
			\vertex at (-2.5,-1.5) (d);
			
			\vertex at (-2.7,-1.7) () {\(S\)};
			\vertex at (-2.7,1.7) () {\(S^\prime\)};
			\vertex at (1.7,1.7) () {\(W^+\)};
			\vertex at (1.7,-1.7) () {\( W^- \)};
			\vertex at (-0.5,-0.35) () {\(Z,  \gamma \)};
			\diagram*{
				(i2) -- [scalar] (i1), (i1) -- [photon] (a), (b) -- [photon] (i1), (i2) -- [scalar] (c),(d) -- [scalar] (i2)
			};
		\end{feynman}
		\node at (-0.6, -2.5) {(g)};
	\end{tikzpicture}
	\hspace{0.3cm}
	\begin{tikzpicture}[scale=0.6]
		\begin{feynman}
			\vertex at (0,0) (a);
			\vertex at (1.7,-1) (i1);
			\vertex at (-1.7,-1) (i2);
			\vertex at (0,1.5) (b);
			\vertex at (1.7,2.5) (c);
			\vertex at (-1.7,2.5) (d);
			\vertex at (1.8,2.7) () {\(W^-(Z)\)};
			\vertex at (-1.8,2.7) () {\(S^\prime\)};
			\vertex at (1.8,-1.3) () {\(W^+(Z)\)};
			\vertex at (-1.8,-1.3) () {\(S\)};
			\vertex at (-1.5,0.7) () {\(H_{1,2}^0, A_{1,2}^0 \)};
			\vertex at (0.7,0.7) () {\((S) \)};
			\diagram*{
				(a) -- [photon] (i1), (i2) -- [scalar] (a),
				(b) -- [scalar] (a), (c) -- [photon] (b), (d) -- [scalar] (b),
			};
		\end{feynman}
		\node at (0, -2.5) {(h)};
	\end{tikzpicture}	
	\hspace{0.3cm}
	\begin{tikzpicture}[scale=0.6]
		\begin{feynman}
			\vertex at (0,0) (i1);
			\vertex at (-1.5,0) (i2);
			\vertex at (1.5,1.5) (a);
			\vertex at (1.5,-1.5) (b);
			\vertex at (-2.5,1.5) (c);
			\vertex at (-2.5,-1.5) (d);
			
			\vertex at (-2.7,-1.7) () {\(S\)};
			\vertex at (-2.7,1.7) () {\(S^\prime\)};
			\vertex at (1.7,1.7) () {\(h\)};
			\vertex at (1.7,-1.7) () {\( h \)};
			\vertex at (-0.8,-0.2) () {\(h\)};
			\diagram*{
				(i2) -- [scalar] (i1), (i1) -- [scalar] (a), (b) -- [scalar] (i1), (i2) -- [scalar] (c),(d) -- [scalar] (i2)
			};
		\end{feynman}
		\node at (-0.6, -2.5) {(i)};
	\end{tikzpicture}
	\hspace{0.3cm}
	\begin{tikzpicture}[scale=0.6]
		\begin{feynman}
			\vertex at (0,0) (a);
			\vertex at (1.7,-1) (i1);
			\vertex at (-1.7,-1) (i2);
			\vertex at (0,1.5) (b);
			\vertex at (1.7,2.5) (c);
			\vertex at (-1.7,2.5) (d);
			\vertex at (1.8,2.7) () {\(h\)};
			\vertex at (-1.8,2.7) () {\(S^\prime\)};
			\vertex at (1.8,-1.3) () {\(h\)};
			\vertex at (-1.8,-1.3) () {\(S\)};
			\vertex at (0.6,1) () {\(S \)};
			
			\diagram*{
				(a) -- [scalar] (i1), (i2) -- [scalar] (a),
				(b) -- [scalar] (a), (c) -- [scalar] (b), (d) -- [scalar] (b),
			};
		\end{feynman}
		\node at (0, -2.5) {(j)};
	\end{tikzpicture}
	\hspace{0.3cm}
	\begin{tikzpicture}[scale=0.6]
		\begin{feynman}
			\vertex at (0,0) (i1);
			\vertex at (2.2,1.7) (a);
			\vertex at (2.2,-1.5) (b);
			\vertex at (-2.2,1.7) (c);
			\vertex at (-2.2,-1.5) (d);
			
			\vertex at (-2.4,-1.7) () {\(S\)};
			\vertex at (-2.4,1.9) () {\(S^\prime \)};
			\vertex at (2.4,1.9) () {\(\gamma\)};
			\vertex at (2.4,-1.7) () {\(\gamma\)};
			\diagram*{
				(i1) -- [photon] (a), (b) -- [photon] (i1), (c) -- [scalar] (i1),(d) -- [scalar] (i1)
			};
		\end{feynman}
		\node at (0, -2.5) {(k)};
	\end{tikzpicture}
			
	\hspace{0.3cm}
	\begin{tikzpicture}[scale=0.6]
		\begin{feynman}
			\vertex at (0,0) (a);
			\vertex at (1.7,-1) (i1);
			\vertex at (-1.7,-1) (i2);
			\vertex at (0,1.5) (b);
			\vertex at (1.7,2.5) (c);
			\vertex at (-1.7,2.5) (d);
			\vertex at (1.8,2.9) () {\(\gamma,Z\)};
			\vertex at (-1.8,2.7) () {\(S^\prime\)};
			\vertex at (1.8,-1.3) () {\(\gamma\)};
			\vertex at (-1.8,-1.3) () {\(S\)};
			\vertex at (0.6,1) () {\(S \)};
			
			\diagram*{
				(a) -- [boson] (i1), (i2) -- [scalar] (a),
				(b) -- [scalar] (a), (c) -- [boson] (b), (d) -- [scalar] (b),
			};
		\end{feynman}
		\node at (0, -2.5) {(l)};
	\end{tikzpicture}
    	\centering
	\begin{tikzpicture}[scale=0.6]
		\begin{feynman}
			\vertex at (0,0) (a);
			\vertex at (1.7,-1) (i1);
			\vertex at (-1.7,-1) (i2);
			\vertex at (0,1.5) (b);
			\vertex at (1.7,2.5) (c);
			\vertex at (-1.7,2.5) (d);
			\vertex at (1.8,2.7) () {\(W^+ \)};
			\vertex at (-1.8,2.7) () {\(S\)};
			\vertex at (1.8,-1.3) () {\(W^+\)};
			\vertex at (-1.8,-1.3) () {\(S\)};
			\vertex at (1.5,1) () {\(H_{1,2}^0 ,A_{1,2}^0\)};
			\diagram*{
				(a) -- [boson] (i1), (i2) -- [scalar] (a),
				(b) -- [scalar] (a), (c) -- [boson] (b), (d) -- [scalar] (b),
			};
		\end{feynman}
		\node at (0, -2.5) {(m)};
	\end{tikzpicture}
	\caption{ Co-annihilation channels for $S^\prime=H_{1,2}^-$ and $S=H_{1,2}^+$, where $f$ represent quark. }
	\label{fig10}
\end{figure}

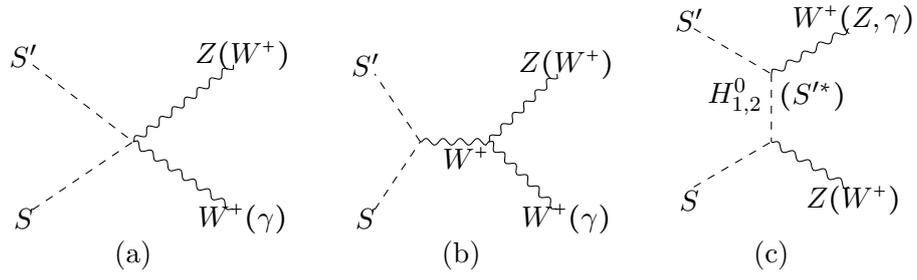
\begin{figure}[hbt!]
	\centering
	\begin{tikzpicture}[scale=0.6]
		\begin{feynman}
			\vertex at (0,0) (i1);
			\vertex at (2.2,1.7) (a);
			\vertex at (2.2,-1.5) (b);
			\vertex at (-2.2,1.7) (c);
			\vertex at (-2.2,-1.5) (d);
			
			\vertex at (-2.4,-1.7) () {\(S\)};
			\vertex at (-2.4,1.9) () {\(S^\prime \)};
			\vertex at (2.4,1.9) () {\(Z (W^+)\)};
			\vertex at (2.4,-1.7) () {\(W^+(\gamma)\)};
			\diagram*{
				(i1) -- [photon] (a), (b) -- [photon] (i1), (c) -- [scalar] (i1),(d) -- [scalar] (i1)
			};
		\end{feynman}
		\node at (0, -2.5) {(a)};
	\end{tikzpicture}
	\hspace{0.3cm}
	\begin{tikzpicture}[scale=0.6]
		\begin{feynman}
			\vertex at (0,0) (i1);
			\vertex at (-1.5,0) (i2);
			\vertex at (1.5,1.5) (a);
			\vertex at (1.5,-1.5) (b);
			\vertex at (-2.5,1.5) (c);
			\vertex at (-2.5,-1.5) (d);
			
			\vertex at (-2.7,-1.7) () {\(S\)};
			\vertex at (-2.7,1.7) () {\(S^\prime\)};
			\vertex at (1.7,1.7) () {\(Z(W^+)\)};
			\vertex at (1.7,-1.7) () {\(W^+(\gamma) \)};
			\vertex at (-0.5,-0.35) () {\(W^+\)};
			\diagram*{
				(i2) -- [boson] (i1), (i1) -- [photon] (a), (b) -- [photon] (i1), (i2) -- [scalar] (c),(d) -- [scalar] (i2)
			};
		\end{feynman}
		\node at (-0.6, -2.5) {(b)};
	\end{tikzpicture}
	\hspace{0.3cm}
	\begin{tikzpicture}[scale=0.6]
		\begin{feynman}
			\vertex at (0,0) (a);
			\vertex at (1.7,-1) (i1);
			\vertex at (-1.7,-1) (i2);
			\vertex at (0,1.5) (b);
			\vertex at (1.7,2.5) (c);
			\vertex at (-1.7,2.5) (d);
			\vertex at (1.8,2.7) () {\(W^+ (Z,\gamma)\)};
			\vertex at (-1.8,2.7) () {\(S^\prime\)};
			\vertex at (1.8,-1.3) () {\(Z(W^+)\)};
			\vertex at (-1.8,-1.3) () {\(S\)};
			\vertex at (-0.8,1) () {\(H_{1,2}^0 \)};
			\vertex at (0.9,1) () {\( ( S^{\prime *}) \)};
			\diagram*{
				(a) -- [boson] (i1), (i2) -- [scalar] (a),
				(b) -- [scalar] (a), (c) -- [boson] (b), (d) -- [scalar] (b),
			};
		\end{feynman}
		\node at (0, -2.5) {(c)};
	\end{tikzpicture}
	\caption{Co-annihilation channels for $S=H^0_{1,2}, A^0_{1,2}$ and $S^\prime=H_{1,2}^+$.}
	\label{fig11}
\end{figure}


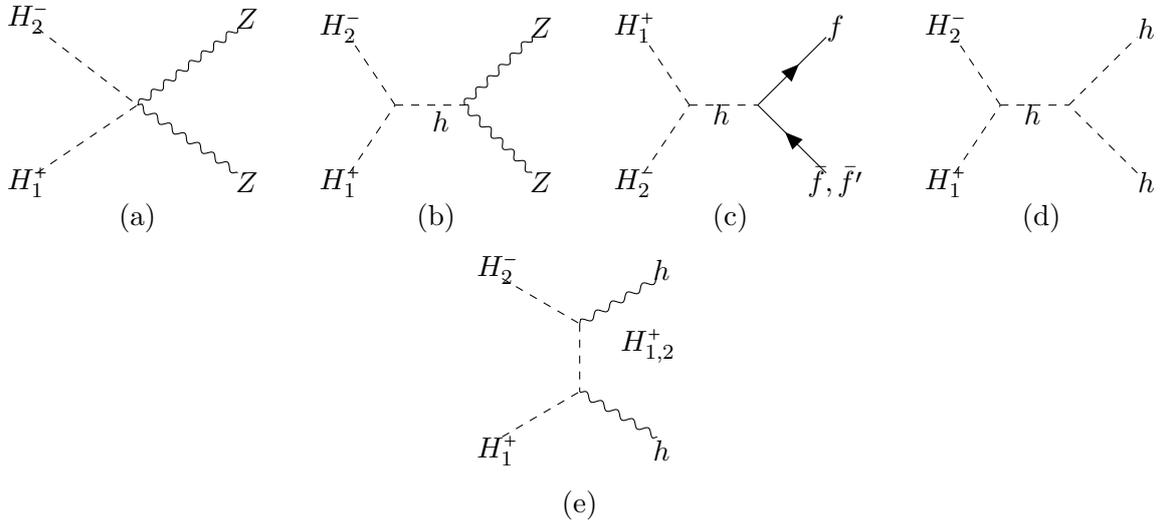
\begin{figure}[hbt!]
	\centering
	\begin{tikzpicture}[scale=0.6]
		\begin{feynman}
			\vertex at (0,0) (i1);
			\vertex at (2.2,1.7) (a);
			\vertex at (2.2,-1.5) (b);
			\vertex at (-2.2,1.7) (c);
			\vertex at (-2.2,-1.5) (d);
			
			\vertex at (-2.4,-1.7) () {\(H_1^+\)};
			\vertex at (-2.4,1.9) () {\(H_2^- \)};
			\vertex at (2.4,1.9) () {\(Z\)};
			\vertex at (2.4,-1.7) () {\(Z\)};
			\diagram*{
				(i1) -- [photon] (a), (b) -- [photon] (i1), (c) -- [scalar] (i1),(d) -- [scalar] (i1)
			};
		\end{feynman}
		\node at (0, -2.5) {(a)};
	\end{tikzpicture}
	\hspace{0.3cm}
	\begin{tikzpicture}[scale=0.6]
		\begin{feynman}
			\vertex at (0,0) (i1);
			\vertex at (-1.5,0) (i2);
			\vertex at (1.5,1.5) (a);
			\vertex at (1.5,-1.5) (b);
			\vertex at (-2.5,1.5) (c);
			\vertex at (-2.5,-1.5) (d);
			
			\vertex at (-2.7,-1.7) () {\(H_1^+\)};
			\vertex at (-2.7,1.7) () {\(H_2^-\)};
			\vertex at (1.7,1.7) () {\(Z\)};
			\vertex at (1.7,-1.7) () {\( Z \)};
			\vertex at (-0.5,-0.35) () {\(h\)};
			\diagram*{
				(i2) -- [scalar] (i1), (i1) -- [photon] (a), (b) -- [photon] (i1), (i2) -- [scalar] (c),(d) -- [scalar] (i2)
			};
		\end{feynman}
		\node at (-0.6, -2.5) {(b)};
	\end{tikzpicture}
	\hspace{0.3cm}
	\begin{tikzpicture}[scale=0.6]
		\begin{feynman}
			\vertex at (0,0) (i1);
			\vertex at (-1.5,0) (i2);
			\vertex at (1.5,1.5) (a);
			\vertex at (1.5,-1.5) (b);
			\vertex at (-2.5,1.5) (c);
			\vertex at (-2.5,-1.5) (d);
			
			\vertex at (-2.7,-1.7) () {\(H_2^-\)};
			\vertex at (-2.7,1.7) () {\(H_1^+\)};
			\vertex at (1.7,1.7) () {\(f\)};
			\vertex at (1.7,-1.7) () {\( \bar{f}, \bar{f'} \)};
			\vertex at (-0.8,-0.2) () {\(h\)};
			\diagram*{
				(i2) -- [scalar] (i1), (i1) -- [fermion] (a), (b) -- [fermion] (i1), (i2) -- [scalar] (c),(d) -- [scalar] (i2)
			};
		\end{feynman}
		\node at (-0.6, -2.5) {(c)};
	\end{tikzpicture}
	\hspace{0.3cm}
	\begin{tikzpicture}[scale=0.6]
		\begin{feynman}
			\vertex at (0,0) (i1);
			\vertex at (-1.5,0) (i2);
			\vertex at (1.5,1.5) (a);
			\vertex at (1.5,-1.5) (b);
			\vertex at (-2.5,1.5) (c);
			\vertex at (-2.5,-1.5) (d);
			
			\vertex at (-2.7,-1.7) () {\(H_1^+\)};
			\vertex at (-2.7,1.7) () {\(H_2^-\)};
			\vertex at (1.7,1.7) () {\(h\)};
			\vertex at (1.7,-1.7) () {\( h \)};
			\vertex at (-0.8,-0.2) () {\(h\)};
			\diagram*{
				(i2) -- [scalar] (i1), (i1) -- [scalar] (a), (b) -- [scalar] (i1), (i2) -- [scalar] (c),(d) -- [scalar] (i2)
			};
		\end{feynman}
		\node at (-0.6, -2.5) {(d)};
	\end{tikzpicture}
	\hspace{0.3cm}
	\begin{tikzpicture}[scale=0.6]
		\begin{feynman}
			\vertex at (0,0) (a);
			\vertex at (1.7,-1) (i1);
			\vertex at (-1.7,-1) (i2);
			\vertex at (0,1.5) (b);
			\vertex at (1.7,2.5) (c);
			\vertex at (-1.7,2.5) (d);
			\vertex at (1.8,2.7) () {\(h \)};
			\vertex at (-1.8,2.7) () {\(H_2^-\)};
			\vertex at (1.8,-1.3) () {\(h\)};
			\vertex at (-1.8,-1.3) () {\(H_1^+\)};
			\vertex at (1.5,1) () {\(H_{1,2}^+\)};
			\diagram*{
				(a) -- [boson] (i1), (i2) -- [scalar] (a),
				(b) -- [scalar] (a), (c) -- [boson] (b), (d) -- [scalar] (b),
			};
		\end{feynman}
		\node at (0, -2.5) {(e)};
	\end{tikzpicture}
	\caption{Dominating co-annihilation channels, where $f$ represent quark. }
	\label{fig13}
\end{figure}


\begin{figure}[hbt!]
	\centering
	\begin{tikzpicture}[scale=0.6]
		\begin{feynman}
			\vertex at (0,0) (i1);
			\vertex at (-1.5,0) (i2);
			\vertex at (1.5,1.5) (a);
			\vertex at (1.5,-1.5) (b);
			\vertex at (-2.5,1.5) (c);
			\vertex at (-2.5,-1.5) (d);
			
			\vertex at (-2.7,-1.7) () {\(\Sigma^+\)};
			\vertex at (-2.7,1.7) () {\(\Sigma^-\)};
			\vertex at (1.7,1.7) () {\(f(l^-)\)};
			\vertex at (1.7,-1.7) () {\( \bar{f}(l^+) \)};
			\vertex at (-0.8,-0.2) () {\(\gamma,Z\)};
			\diagram*{
				(i2) -- [photon] (i1), (i1) -- [fermion] (a), (b) -- [fermion] (i1), (i2) -- [fermion] (c),(d) -- [fermion] (i2)
			};
		\end{feynman}
		\node at (-0.5, -2.5) {(a)};
	\end{tikzpicture}
	\hspace{0.3cm}
	\begin{tikzpicture}[scale=0.6]
		\begin{feynman}
			\vertex at (0,0) (a);
			\vertex at (1.7,-1) (i1);
			\vertex at (-1.7,-1) (i2);
			\vertex at (0,1.5) (b);
			\vertex at (1.7,2.5) (c);
			\vertex at (-1.7,2.5) (d);
			\vertex at (1.8,2.7) () {\(l^-\)};
			\vertex at (-1.8,2.7) () {\(\Sigma^- \)};
			\vertex at (1.8,-1.3) () {\(l^+\)};
			\vertex at (-1.8,-1.3) () {\(\Sigma^+\)};
			\vertex at (1.5,1) () {\( H_{1,2}^0,A^0_{1,2} \)};
			
			\diagram*{
				(i1) -- [fermion] (a), (i2) -- [fermion] (a),
				(b) -- [scalar] (a), (b) -- [fermion] (c), (d) -- [fermion] (b),
			};
		\end{feynman}
		\node at (0, -2.5) {(b)};
	\end{tikzpicture}
	\hspace{0.3cm}
	\begin{tikzpicture}[scale=0.6]
		\begin{feynman}
			\vertex at (0,0) (i1);
			\vertex at (-1.5,0) (i2);
			\vertex at (1.5,1.5) (a);
			\vertex at (1.5,-1.5) (b);
			\vertex at (-2.5,1.5) (c);
			\vertex at (-2.5,-1.5) (d);
			
			\vertex at (-2.7,-1.7) () {\(\Sigma^+\)};
			\vertex at (-2.7,1.7) () {\(\Sigma^-\)};
			\vertex at (1.7,1.7) () {\(\nu\)};
			\vertex at (1.7,-1.7) () {\( \nu \)};
			\vertex at (-0.8,-0.2) () {\(Z\)};
			\diagram*{
				(i1) -- [photon] (i2), (i1) -- [fermion] (a), (i1) -- [fermion] (b), (i2) -- [fermion] (c),(d) -- [fermion] (i2)
			};
		\end{feynman}
		\node at (-0.5, -2.5) {(c)};
	\end{tikzpicture}
	\hspace{0.3cm}
	\begin{tikzpicture}[scale=0.6]
		\begin{feynman}
			\vertex at (0,0) (a); 
			\vertex at (1.7,-1) (i1); 
			\vertex at (-1.7,-1) (i2); 
			\vertex at (0,1.5) (b); 
			\vertex at (1.7,2.5) (c); 
			\vertex at (-1.7,2.5) (d); 
			\vertex at (1.8,2.7) () {\(\nu\)};
			\vertex at (-1.8,2.7) () {\(\Sigma^- \)};
			\vertex at (1.8,-1.3) () {\(\nu,\nu'\)};
			\vertex at (-1.8,-1.3) () {\(\Sigma^+\)};
			\vertex at (0.7,1) () {\( H_{1,2}^+ \)};
			
			\diagram*{
				(a) -- [fermion] (i1), (i2) -- [fermion] (a),
				(a) -- [scalar] (b), (b) -- [fermion] (c), (d) -- [fermion] (b),
			};
		\end{feynman}
		\node at (0, -2.5) {(d)};
	\end{tikzpicture}
\hspace{0.3cm}
\begin{tikzpicture}[scale=0.6]
	\begin{feynman}
		\vertex at (0,0) (a);
		\vertex at (1.7,-1) (i1);
		\vertex at (-1.7,-1) (i2);
		\vertex at (0,1.5) (b);
		\vertex at (1.7,2.5) (c);
		\vertex at (-1.7,2.5) (d);
		\vertex at (1.8,2.7) () {\(\gamma,Z\)};
		\vertex at (-1.8,2.7) () {\(\Sigma^-\)};
		\vertex at (1.8,-1.3) () {\(\gamma,Z\)};
		\vertex at (-1.8,-1.3) () {\(\Sigma^+\)};
		\vertex at (0.4,1) () {\( \Sigma^+ \)};
		
		\diagram*{
			(i1) -- [photon] (a), (i2) -- [fermion] (a),
			(a) -- [fermion] (b), (b) -- [photon] (c), (b) -- [fermion] (d),
		};
	\end{feynman}
	\node at (0, -2.5) {(e)};
\end{tikzpicture}
	\hspace{0.3cm}
	\begin{tikzpicture}[scale=0.6]
		\begin{feynman}
			\vertex at (0,0) (i1);
			\vertex at (-1.5,0) (i2);
			\vertex at (1.5,1.5) (a);
			\vertex at (1.5,-1.5) (b);
			\vertex at (-2.5,1.5) (c);
			\vertex at (-2.5,-1.5) (d);
			
			\vertex at (-2.7,-1.7) () {\(\Sigma^+\)};
			\vertex at (-2.7,1.7) () {\(\Sigma^-\)};
			\vertex at (1.7,1.7) () {\(W^+\)};
			\vertex at (1.7,-1.7) () {\( W^- \)};
			\vertex at (-0.8,-0.2) () {\(\gamma,Z\)};
			\diagram*{
				(i2) -- [photon] (i1), (i1) -- [boson] (a), (b) -- [boson] (i1), (i2) -- [fermion] (c),(d) -- [fermion] (i2)
			};
		\end{feynman}
		\node at (-0.5, -2.5) {(f)};
	\end{tikzpicture}
	\hspace{0.3cm}
	\begin{tikzpicture}[scale=0.6]
		\begin{feynman}
			\vertex at (0,0) (a);
			\vertex at (1.7,-1) (i1);
			\vertex at (-1.7,-1) (i2);
			\vertex at (0,1.5) (b);
			\vertex at (1.7,2.5) (c);
			\vertex at (-1.7,2.5) (d);
			\vertex at (1.8,2.7) () {\(W^-\)};
			\vertex at (-1.8,2.7) () {\(\Sigma^-\)};
			\vertex at (1.8,-1.3) () {\(W^+\)};
			\vertex at (-1.8,-1.3) () {\(\Sigma^+\)};
			\vertex at (0.4,1) () {\( \Sigma^0 \)};
			
			\diagram*{
				(i1) -- [photon] (a), (i2) -- [fermion] (a),
				(a) -- [fermion] (b), (b) -- [photon] (c), (b) -- [fermion] (d),
			};
		\end{feynman}
		\node at (0, -2.5) {(g)};
	\end{tikzpicture}
	\hspace{0.3cm}
	\begin{tikzpicture}[scale=0.6]
		\begin{feynman}
			\vertex at (0,0) (a);
			\vertex at (1.7,-1) (i1);
			\vertex at (-1.7,-1) (i2);
			\vertex at (0,1.5) (b);
			\vertex at (1.7,2.5) (c);
			\vertex at (-1.7,2.5) (d);
			\vertex at (1.8,2.7) () {\(W^+\)};
			\vertex at (-1.8,2.7) () {\(\Sigma^+\)};
			\vertex at (1.8,-1.3) () {\(W^+\)};
			\vertex at (-1.8,-1.3) () {\(\Sigma^+\)};
			\vertex at (0.4,1) () {\( \Sigma^0 \)};
			
			\diagram*{
				(i1) -- [photon] (a), (i2) -- [fermion] (a),
				(a) -- [fermion] (b), (b) -- [photon] (c), (b) -- [fermion] (d),
			};
		\end{feynman}
		\node at (0, -2.5) {(h)};
	\end{tikzpicture}
	\hspace{0.3cm}
	\begin{tikzpicture}[scale=0.6]
		\begin{feynman}
			\vertex at (0,0) (a);
			\vertex at (1.7,-1) (i1);
			\vertex at (-1.7,-1) (i2);
			\vertex at (0,1.5) (b);
			\vertex at (1.7,2.5) (c);
			\vertex at (-1.7,2.5) (d);
			\vertex at (1.8,2.7) () {\(W^-(W^+)\)};
			\vertex at (-1.8,2.7) () {\(\Sigma^0\)};
			\vertex at (1.8,-1.3) () {\(W^+(W^-)\)};
			\vertex at (-1.8,-1.3) () {\(\Sigma^0\)};
			\vertex at (1.2,1) () {\( \Sigma^-(\Sigma^+) \)};
			
			\diagram*{
				(i1) -- [photon] (a), (a) -- [fermion] (i2),
				(b) -- [fermion] (a), (b) -- [photon] (c), (d) -- [fermion] (b),
			};
		\end{feynman}
		\node at (0, -2.5) {(i)};
	\end{tikzpicture}
	\centering
	\begin{tikzpicture}[scale=0.6]
		\begin{feynman}
			\vertex at (0,0) (i1); 
			\vertex at (-1.5,0) (i2); 
			\vertex at (1.5,1.5) (a); 
			\vertex at (1.5,-1.5) (b); 
			\vertex at (-2.5,1.5) (c); 
			\vertex at (-2.5,-1.5) (d); 
			
			\vertex at (-2.7,-1.7) () {\(\Sigma^0\)};
			\vertex at (-2.7,1.7) () {\(\Sigma^+\)};
			\vertex at (1.7,1.7) () {\(f\)};
			\vertex at (1.7,-1.7) () {\( \bar{f'}\)};
			\vertex at (-0.8,-0.2) () {\(W^+\)};
			\diagram*{
				(i2) -- [photon] (i1), (i1) -- [fermion] (a), (b) -- [fermion] (i1), (c) -- [fermion] (i2),(d) -- [fermion] (i2)
			};
		\end{feynman}
		\node at (-0.5, -2.5) {(j)};
	\end{tikzpicture}
			%
		\hspace{0.3cm}
		\begin{tikzpicture}[scale=0.6]
			\begin{feynman}
				\vertex at (0,0) (i1); 
				\vertex at (-1.5,0) (i2); 
				\vertex at (1.5,1.5) (a); 
				\vertex at (1.5,-1.5) (b); 
				\vertex at (-2.5,1.5) (c); 
				\vertex at (-2.5,-1.5) (d); 
				
				\vertex at (-2.7,-1.7) () {\(\Sigma^0\)};
				\vertex at (-2.7,1.7) () {\(\Sigma^+\)};
				\vertex at (1.7,1.7) () {\(W^+\)};
				\vertex at (1.7,-1.7) () {\( \gamma\)};
				\vertex at (-0.8,-0.2) () {\(W^+\)};
				\diagram*{
					(i2) -- [photon] (i1), (i1) -- [photon] (a), (b) -- [photon] (i1), (c) -- [fermion] (i2),(d) -- [fermion] (i2)
				};
			\end{feynman}
			\node at (-0.5, -2.5) {(k)};
		\end{tikzpicture}
		\hspace{0.3cm}
		\begin{tikzpicture}[scale=0.6]
			\begin{feynman}
				\vertex at (0,0) (a); 
				\vertex at (1.7,-1) (i1); 
				\vertex at (-1.7,-1) (i2); 
				\vertex at (0,1.5) (b); 
				\vertex at (1.7,2.5) (c); 
				\vertex at (-1.7,2.5) (d); 
				\vertex at (1.8,2.7) () {\(\gamma\)};
				\vertex at (-1.8,2.7) () {\(\Sigma^+ \)};
				\vertex at (1.8,-1.3) () {\(W^+\)};
				\vertex at (-1.8,-1.3) () {\(\Sigma^0\)};
				\vertex at (0.4,1) () {\( \Sigma^- \)};
				
				\diagram*{
					(a) -- [photon] (i1), (i2) -- [fermion] (a),
					(b) -- [fermion] (a), (b) -- [photon] (c), (d) -- [fermion] (b),
				};
			\end{feynman}
			\node at (0, -2.5) {(l)};
		\end{tikzpicture}
		\hspace{0.3cm}
		\begin{tikzpicture}[scale=0.6]
			\begin{feynman}
				\vertex at (0,0) (i1); 
				\vertex at (-1.5,0) (i2); 
				\vertex at (1.5,1.5) (a); 
				\vertex at (1.5,-1.5) (b); 
				\vertex at (-2.5,1.5) (c); 
				\vertex at (-2.5,-1.5) (d); 
				
				\vertex at (-2.7,-1.7) () {\(\Sigma^0\)};
				\vertex at (-2.7,1.7) () {\(\Sigma^+\)};
				\vertex at (1.7,1.7) () {\(Z\)};
				\vertex at (1.7,-1.7) () {\( W^+\)};
				\vertex at (-0.8,-0.2) () {\(W^+\)};
				\diagram*{
					(i2) -- [photon] (i1), (i1) -- [photon] (a), (b) -- [photon] (i1), (c) -- [fermion] (i2),(d) -- [fermion] (i2)
				};
			\end{feynman}
			\node at (-0.5, -2.5) {(m)};
		\end{tikzpicture}
		\hspace{0.3cm}
		\begin{tikzpicture}[scale=0.6]
			\begin{feynman}
				\vertex at (0,0) (a); 
				\vertex at (1.7,-1) (i1); 
				\vertex at (-1.7,-1) (i2); 
				\vertex at (0,1.5) (b); 
				\vertex at (1.7,2.5) (c); 
				\vertex at (-1.7,2.5) (d); 
				\vertex at (1.8,2.7) () {\(Z\)};
				\vertex at (-1.8,2.7) () {\(\Sigma^+ \)};
				\vertex at (1.8,-1.3) () {\(W^+\)};
				\vertex at (-1.8,-1.3) () {\(\Sigma^0\)};
				\vertex at (0.4,1) () {\( \Sigma^- \)};
				
				\diagram*{
					(a) -- [photon] (i1), (i2) -- [fermion] (a),
					(b) -- [fermion] (a), (b) -- [photon] (c), (d) -- [fermion] (b),
				};
			\end{feynman}
			\node at (0, -2.5) {(n)};
		\end{tikzpicture}
    \hspace{0.3cm}
		\begin{tikzpicture}[scale=0.6]
		\begin{feynman}
			\vertex at (0,0) (i1);
			\vertex at (-1.5,0) (i2);
			\vertex at (1.5,1.5) (a);
			\vertex at (1.5,-1.5) (b);
			\vertex at (-2.5,1.5) (c);
			\vertex at (-2.5,-1.5) (d);
			
			\vertex at (-2.7,-1.7) () {\(\Sigma^0\)};
			\vertex at (-2.7,1.7) () {\(\Sigma^+\)};
			\vertex at (1.7,1.7) () {\(W^+\)};
			\vertex at (1.7,-1.7) () {\( h \)};
			\vertex at (-0.8,-0.2) () {\(W^+\)};
			\diagram*{
				(i2) -- [photon] (i1), (i1) -- [boson] (a), (b) -- [scalar] (i1), (c) -- [fermion] (i2),(d) -- [fermion] (i2)
			};
		\end{feynman}
		\node at (-0.5, -2.5) {(o)};
	\end{tikzpicture}
	\caption{Dominating (co-)annihilation channels where $f$ ($l$) represent  quark (lepton).}
	\label{fig14}
\end{figure}
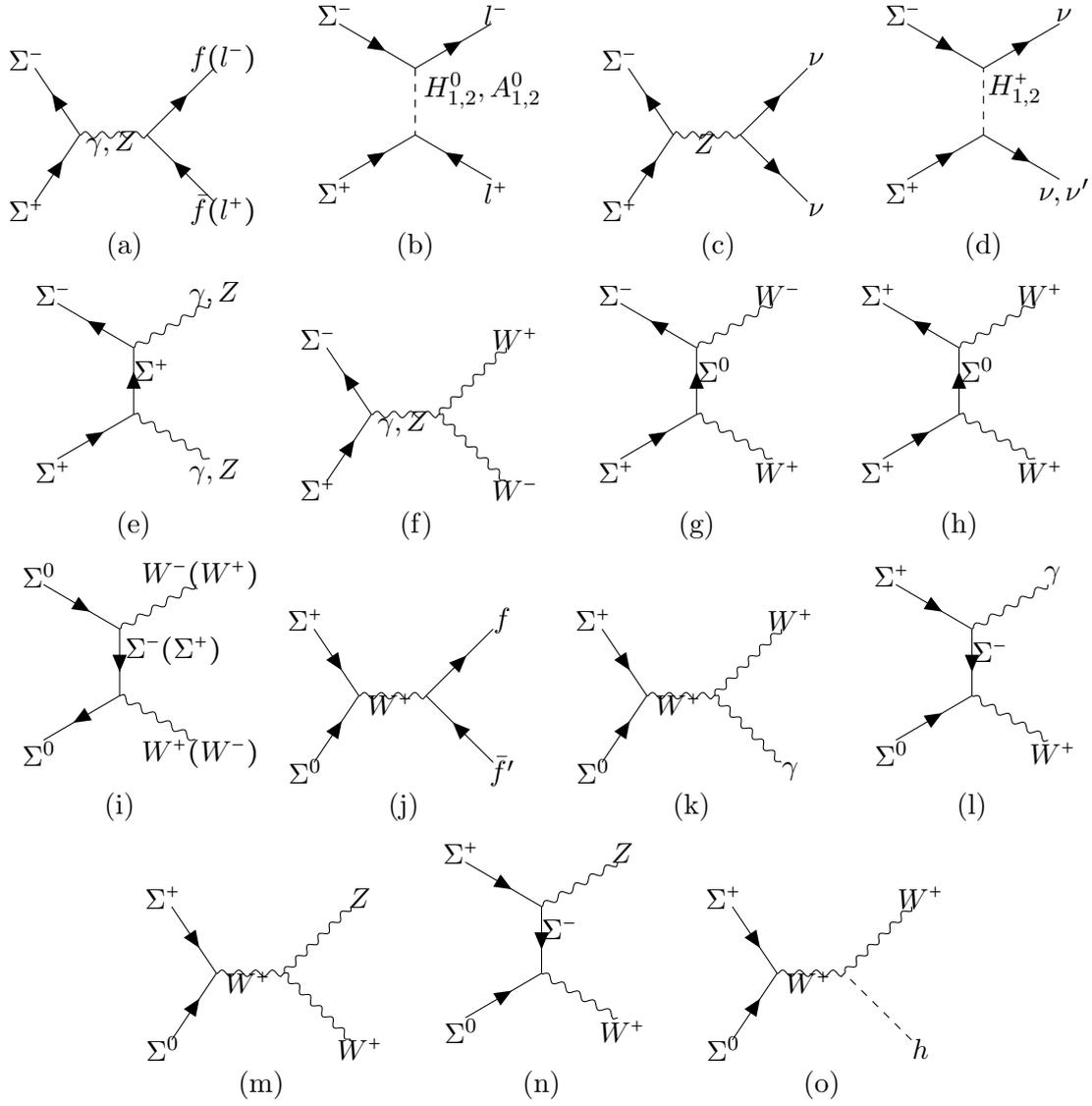

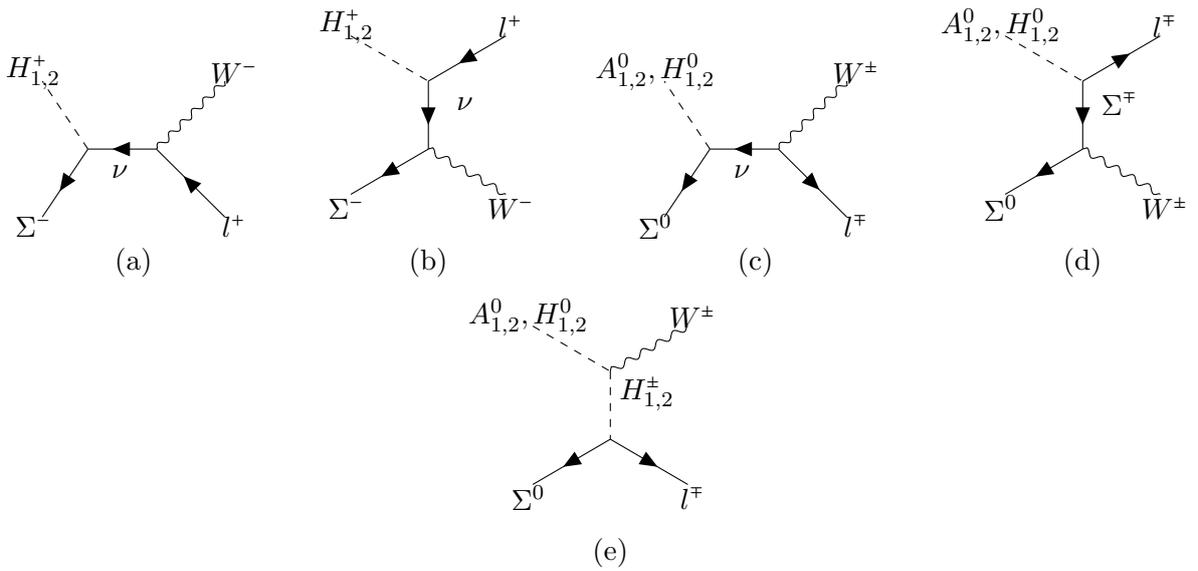
\begin{figure}[hbt!]
	\centering
	\begin{tikzpicture}[scale=0.6]
		\begin{feynman}
			\vertex at (0,0) (i1);
			\vertex at (-1.5,0) (i2);
			\vertex at (1.5,1.5) (a);
			\vertex at (1.5,-1.5) (b);
			\vertex at (-2.5,1.5) (c);
			\vertex at (-2.5,-1.5) (d);
			
			\vertex at (-2.7,-1.7) () {\(\Sigma^-\)};
			\vertex at (-2.7,1.7) () {\(H_{1,2}^+\)};
			\vertex at (1.7,1.7) () {\(W^-\)};
			\vertex at (1.7,-1.7) () {\( l^+ \)};
			\vertex at (-0.8,-0.5) () {\(\nu\)};
			\diagram*{
				(i1) -- [fermion] (i2), (i1) -- [boson] (a), (b) -- [fermion] (i1), (i2) -- [scalar] (c),(i2) -- [fermion] (d)
			};
		\end{feynman}
		\node at (-0.5, -2.5) {(a)};
	\end{tikzpicture}
	\hspace{0.3cm}
	\begin{tikzpicture}[scale=0.6]
		\begin{feynman}
			\vertex at (0,0) (a);
			\vertex at (1.7,-1) (i1);
			\vertex at (-1.7,-1) (i2);
			\vertex at (0,1.5) (b);
			\vertex at (1.7,2.5) (c);
			\vertex at (-1.7,2.5) (d);
			\vertex at (1.8,2.7) () {\(l^+\)};
			\vertex at (-1.8,2.7) () {\(H_{1,2}^+\)};
			\vertex at (1.8,-1.3) () {\(W^-\)};
			\vertex at (-1.8,-1.3) () {\(\Sigma^- \)};
			\vertex at (0.8,1) () {\(\nu \)};
			
			\diagram*{
				(i1) -- [boson] (a), (a) -- [fermion] (i2),
				(b) -- [fermion] (a), (c) -- [fermion] (b), (d) -- [scalar] (b),
			};
		\end{feynman}
		\node at (0, -2.5) {(b)};
	\end{tikzpicture}
    \hspace{0.3cm}
    \centering
	\begin{tikzpicture}[scale=0.6]
		\begin{feynman}
			\vertex at (0,0) (i1);
			\vertex at (-1.5,0) (i2);
			\vertex at (1.5,1.5) (a);
			\vertex at (1.5,-1.5) (b);
			\vertex at (-2.5,1.5) (c);
			\vertex at (-2.5,-1.5) (d);
			
			\vertex at (-2.7,-1.7) () {\(\Sigma^0\)};
			\vertex at (-2.7,1.7) () {\(A_{1,2}^0, H_{1,2}^0\)};
			\vertex at (1.7,1.7) () {\(W^\pm\)};
			\vertex at (1.7,-1.7) () {\( l^\mp \)};
			\vertex at (-0.8,-0.5) () {\(\nu\)};
			\diagram*{
				(i1) -- [fermion] (i2), (i1) -- [boson] (a), (i1) -- [fermion] (b), (i2) -- [scalar] (c),(i2) -- [fermion] (d)
			};
		\end{feynman}
		\node at (-0.5, -2.5) {(c)};
	\end{tikzpicture}
    \hspace{0.3cm}
	\begin{tikzpicture}[scale=0.6]
		\begin{feynman}
			\vertex at (0,0) (a);
			\vertex at (1.7,-1) (i1);
			\vertex at (-1.7,-1) (i2);
			\vertex at (0,1.5) (b);
			\vertex at (1.7,2.5) (c);
			\vertex at (-1.7,2.5) (d);
			\vertex at (1.8,2.7) () {\(l^\mp\)};
			\vertex at (-1.8,2.7) () {\(A_{1,2}^0, H_{1,2}^0\)};
			\vertex at (1.8,-1.3) () {\(W^\pm\)};
			\vertex at (-1.8,-1.3) () {\(\Sigma^0 \)};
			\vertex at (0.8,1) () {\(\Sigma^\mp \)};
			
			\diagram*{
				(i1) -- [boson] (a), (a) -- [fermion] (i2),
				(b) -- [fermion] (a), (b) -- [fermion] (c), (d) -- [scalar] (b),
			};
		\end{feynman}
		\node at (0, -2.5) {(d)};
	\end{tikzpicture}
    \hspace{0.3cm}
	\begin{tikzpicture}[scale=0.6]
		\begin{feynman}
			\vertex at (0,0) (a);
			\vertex at (1.7,-1) (i1);
			\vertex at (-1.7,-1) (i2);
			\vertex at (0,1.5) (b);
			\vertex at (1.7,2.5) (c);
			\vertex at (-1.7,2.5) (d);
			\vertex at (1.8,2.7) () {\(W^\pm\)};
			\vertex at (-1.8,2.7) () {\(A_{1,2}^0, H_{1,2}^0\)};
			\vertex at (1.8,-1.3) () {\(l^\mp\)};
			\vertex at (-1.8,-1.3) () {\(\Sigma^0 \)};
			\vertex at (0.8,1) () {\(H_{1,2}^\pm \)};
			
			\diagram*{
				(a) -- [fermion] (i1), (a) -- [fermion] (i2),
				(b) -- [scalar] (a), (c) -- [boson] (b), (d) -- [scalar] (b),
			};
		\end{feynman}
		\node at (0, -2.5) {(e)};
	\end{tikzpicture}
	\caption{Dominating co-annihilation channels of triplet fermion with inert scalars where $l$  represent lepton.}
	\label{fig18}
\end{figure}
\end{appendices}

\end{document}